\documentclass[12pt]{iopart}

\usepackage{graphicx}
\usepackage{hyperref}
\usepackage{verbatim}
\usepackage{array}
\usepackage[table]{xcolor}
\usepackage{multirow}
\usepackage[section]{placeins} 
\usepackage[square,numbers,comma,sort&compress]{natbib} 

\begin{document}

\title[Estimating the NEMA characteristics of the J-PET tomograph using the GATE package]
{Estimating the NEMA characteristics of the J-PET tomograph using the GATE package}

\author{
P.~Kowalski$^1$,
W.~Wi\'slicki$^1$,
R.~Y.~Shopa$^1$,
L.~Raczy\'nski$^1$,
K.~Klimaszewski$^1$,
C.~Curcenau$^3$,
E.~Czerwi\'nski$^2$,
K.~Dulski$^2$,
A.~Gajos$^2$,
M.~Gorgol$^4$,
N.~Gupta-Sharma$^2$,
B.~Hiesmayr$^5$,
B.~Jasi\'nska$^4$,
\L .~Kap\l on$^2$,
D.~Kisielewska-Kami\'nska$^2$,
G.~Korcyl$^2$,
T.~Kozik$^2$,
W.~Krzemie\'n$^6$,
E.~Kubicz$^2$,
M.~Mohammed$^{2,7}$,
S.~Nied\'zwiecki$^2$,
M.~Pa\l ka$^2$,
M.~Pawlik-Nied\'zwiecka$^2$,
J.~Raj$^2$,
K.~Rakoczy$^2$,
Z.~Rudy$^2$,
S.~Sharma$^2$,
S.~Shivani$^2$,
M.~Silarski$^2$,
M.~Skurzok$^2$,
B.~Zgardzi\'nska$^4$,
M.~Zieli\'nski$^2$,
P.~Moskal$^2$
}

\address{$^1$ Department of Complex Systems, National Centre for Nuclear Research, 05-400 Otwock-\'Swierk, Poland}
\address{$^2$ Faculty of Physics, Astronomy and Applied Computer Science, Jagiellonian University, 30-348 Cracow, Poland}
\address{$^3$ INFN, Laboratori Nazionali di Frascati, 00044 Frascati, Italy}
\address{$^4$ Institute of Physics, Maria Curie-Sk\l odowska University, 20-031 Lublin, Poland}
\address{$^5$ Faculty of Physics, University of Vienna, 1090 Vienna, Austria}
\address{$^6$ High Energy Physics Division, National Centre for Nuclear Research, 05-400 Otwock-\'Swierk, Poland}
\address{$^7$ Department of Physics, College of Education for Pure Sciences, University of Mosul, Mosul, Iraq}

\begin{abstract}

The novel whole-body PET system based on plastic scintillators is developed by the {J-PET} Collaboration.
It consists of plastic scintillator strips arranged axially in the form of a~cylinder, allowing the cost-effective construction of the total-body PET.
In order to determine properties of the scanner prototype and optimize its geometry, advanced computer simulations using the GATE software were performed.

The spatial resolution, the sensitivity, the scatter fraction and the noise equivalent count rate were estimated according to the NEMA norm as a~function of the length of the tomograph, number of the detection layers, diameter of the tomographic chamber and for various types of the applied readout.
For the single-layer geometry with the diameter of 85~cm, strip length of 100~cm, cross-section of 4~mm~x~20~mm and silicon photomultipliers with the additional layer of wavelength shifter as the readout, the spatial resolution (FWHM) in the centre of the scanner is equal to 3~mm (radial, tangential) and 6~mm (axial).
For the analogous double-layer geometry with the same readout, diameter and scintillator length, with the strip cross-section of 7~mm~x~20~mm, the NECR peak of 300~kcps was reached at 40~kBq/cc activity concentration, the scatter fraction is estimated to about 35\% and the sensitivity at the centre amounts to 14.9~cps/kBq.
Sensitivity profiles were also determined.

\end{abstract}
\vspace{2pc}
{\noindent{\it Keywords}: NEMA norms, J-PET, positron emission tomography, plastic scintillators}
\submitto{\PMB}
\maketitle

\section{Introduction}

The National Electrical Manufacturers Association (NEMA) is the association of electrical equipment and medical imaging manufacturers in the United States, which publishes standards for medical diagnostic imaging equipment.
One of its standards is NEMA-NU-2 \cite{NEMA}, for the Positron Emission Tomography (PET) devices.
It defines comprehensive characteristics of the PET scanners: the spatial resolution, scatter fraction, noise equivalent count rate (NECR) \cite{Yang2015}, count losses and sensitivity.
They allow to compare different PET tomographs.

The subject of this article is the NEMA characteristics of the PET scanner with a~large Axial Field of View (AFOV) and dependence of these characteristics on the geometry of such a~scanner.
Examples of scanners with large AFOV are: 3D PET scanner based on lead-walled straw detectors \cite{Lacy2010}, RPC-PET based on resistive plate chambers \cite{Crespo2013}, the first generation of the presently developed total-body EXPLORER PET \cite{Viswanath2017, Zhang2017, Cherry2017, Cherry2018} and the Jagiellonian PET (J-PET) scanner based on the plastic scintillator strips \cite{Niedzwiecki2017}.
In this article we present the simulated NEMA characteristics of the J-PET.

The J-PET detector is a~novel PET scanner built from axially arranged plastic scintillator strips \cite{Moskal2011, Moskal2014a, Moskal2015, Moskal2016}, in contrast to the classical PET scanners, based on inorganic crystal scintillators \cite{Moses1999, Moses2003, Humm2003, Townsend2004, Karp2008, Conti2009, Conti2011, Slomka2016}.
In case of crystal tomographs, annihilation photons are registered using photoelectric effect, while in the J-PET the Compton scattering is used.
In the J-PET, relatively low detection efficiency (in comparison to crystal PET scanners) is compensated with a~high time resolution of plastic scintillators, an application of few concentric detection layers, and a~large AFOV of the detector \cite{Moskal2016}.
The use of plastic detectors opens the perspective for the cost-effective construction of the total-body PET.
In such a~total-body PET scanner, it would be possible to image a~whole patient's body without moving the patient along the axis of the device.
The total-body technique would increase the effective sensitivity, decrease the time of the examination and reduce a~necessary image blur due to the patient's or scanner's movements when the whole body has to be examined.
The clinical advantages of the total-body scanner are extensively discussed in \cite{Viswanath2017, Zhang2017, Cherry2017, Cherry2018}.
For completeness we emphasize here a~possibility of the significant reduction of radiation dose needed for the whole-body scan, possibility of usage of shorter lived tracers and possibility of the studies of kinetics in many organs simultaneously.
Additionally, the dynamic range of the total-body PET scanners is much broader than of classical tomographs.
It results in the fact that the radiotracers may be followed for a~longer time before the signal decays to such an extent that it is not detectable.
Also the viability of the multitracer studies is improved.
Building of the total-body PET scanner based on the crystal scintillators is expensive \footnote{
The first estimations of the commercial costs of such scanners points to about 10 mln dollars \cite{Cherry2018}.
}.

Much lower costs of building the J-PET scanner based on plastic scintillators (due to the less expensive detector material and reduced number of the electronic channels) can make the total-body PET diagnostics more widely available.
In addition, in the J-PET solution with axially arranged plastic strips, the readout is placed outside of the detection chamber simplifying PET/MR hybrid construction and enabling extension of the AFOV without a~significant increase of costs.
Moreover, the J-PET solution enables to built a~detector from multi-strip modules, the number of which can be adopted to the size of the patient.
Such modular J-PET can contribute to broader applications of cancer diagnostics, especially for people with large size of the body and claustrophobic disease.
A~first prototype of such portable and modular J-PET with 50~cm AFOV is in the final stage of construction at the Jagiellonian University.
This prototype consists of 24 modules, each with the weight of about 2~kg only \cite{Moskal2018_totalbody}.

The aim of this article is to present the NEMA characteristics of the J-PET scanner as a~function of the length of the tomograph, the number of the detection layers, the diameter of the tomographic chamber and the type of the readout.
The above characteristics may be used as a~figure of merit in the geometry optimization of the prototype device.
Because of the fact that the J-PET scanner is meant to be used in medical diagnostics, the target performance should be at least comparable to the performance of currently available commercial PET devices.
In section~\ref{principleofoperation} the principle of operation of the J-PET scanner is described.
Section~\ref{materialsandmethods} defines the parameters of the tomograph considered in this article (\ref{simulations}), description of the data selection method, which was used to minimize the background from the annihilation photon scattered in the examined object or in the detector (\ref{eventselection}) and details about calculated characteristics (\ref{materials_sensitivity}-\ref{materials_necr}).
Next sections provide the obtained results (\ref{section_reults}) and the summary (\ref{section_summary}).


\section{Principle of operation of the J-PET tomograph}
\label{principleofoperation}

\begin{figure}[!htb]
\begin{center}
\includegraphics[width=0.49\textwidth]{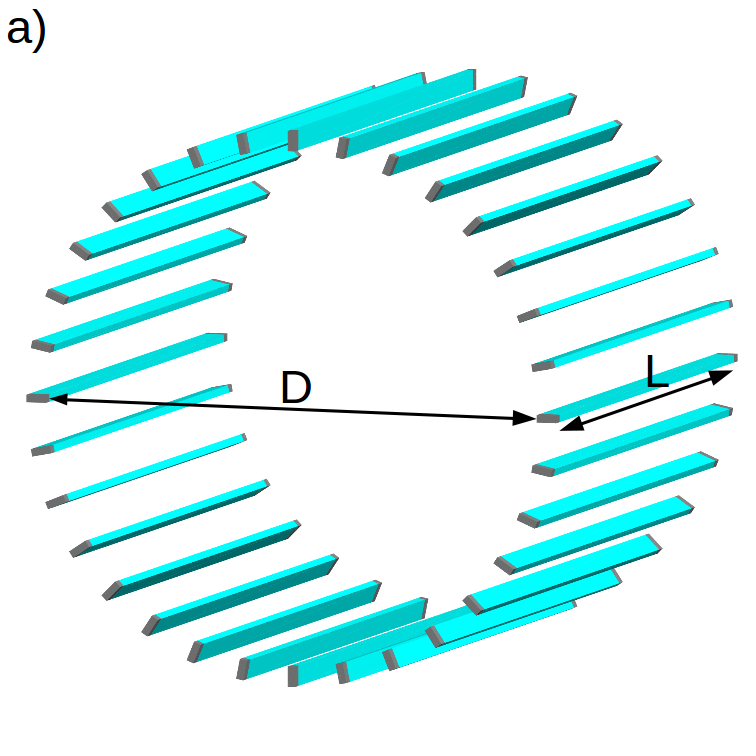}
\includegraphics[width=0.49\textwidth]{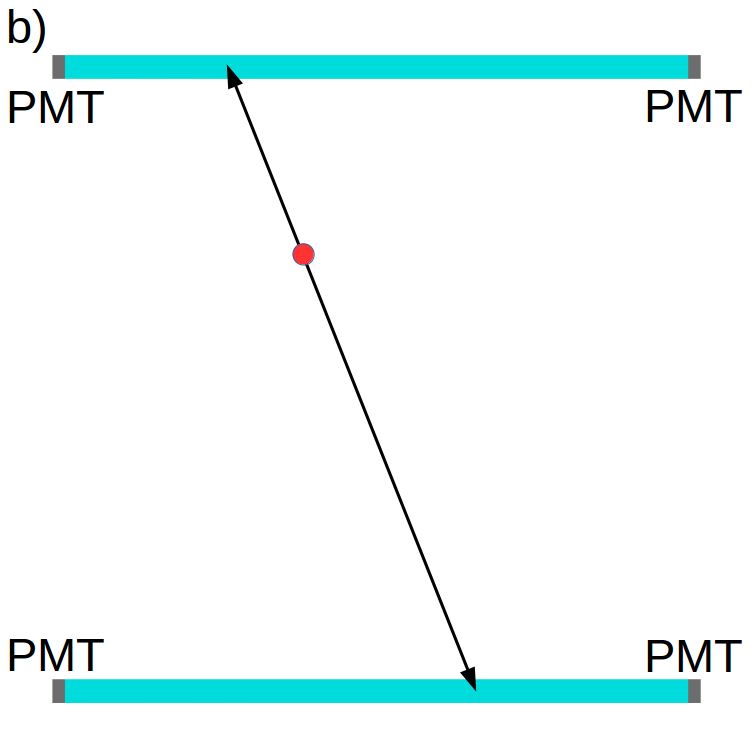}
\end{center}
\caption{
a) The exemplary geometry of the J-PET scanner; scintillators are cyan and photomultipliers (PMT) are gray.
\textbf{D} stands for the diameter of the scanner and \textbf{L} the length.
For better visualization only 32 strips are shown but in our studies the full angle coverage (without gaps) is used.
b) Pair of scintillating strips and two annihilation photons defining the line of response (details in the text). Red point denotes the annihilation point.
\label{geometry_32strips}
}
\end{figure}

The J-PET scanner is built from axially-arranged plastic scintillator strips.
Each scintillator is readout by two photomultipliers located at the ends of the strip.
Optical photons, generated in the scintillator due to the interaction of the gamma photon with the scintillator, propagate to the ends of the strip.
Thus the scintillator strip acts also as a~lightguide.
In the photomultipliers, optical photons are converted into voltage signals and processed using dedicated front-end electronics boards \cite{Palka2017} and the triggerless data acquisition system \cite{Korcyl2016, Korcyl2018}.
Data is stored locally and analyzed using the JPetFramework software \cite{Krzemien2015}.
Exemplary geometrical configuration of a~single-layer J-PET tomograph is visualized in Fig.~\ref{geometry_32strips}a.

Like in classical PET scanners, the J-PET detector registers back-to-back annihilation photons with energy 511~keV, outgoing from the investigated object.
These photons come from annihilations of positrons emitted from the radionuclide (for example $^{18}$F) with electrons in surrounding tissue.

The principle of operation of the J-PET may be explained using a~simplified 2-strip model (Fig.~\ref{geometry_32strips}b).
Having times of arrivals of light signals to the ends of the strips, one can estimate the position of the interaction of the 511~keV photon along the strip and the time of the scattering \cite{Moskal2014a, Moskal2015, Moskal2016}.
In order to calculate these values, specialized algorithms were developed.
In one of these algorithms, voltage signals at photomultipliers were reconstructed using the compressive sensing theory and using these signals, times and positions of interactions were calculated \cite{Raczynski2014, Raczynski2015}.

Experiments showed that 80~ps hit-time resolution (standard deviation) for a~30~cm strip is achievable, allowing to determine the position of the interaction along the strip with spatial resolution of 2.2~cm (FWHM).
The fractional energy resolution for the energy deposited by the annihilation photon in the plastic scintillator strips was measured for a~single strip J-PET prototype and it amounts to $\sigma(E)/E \approx 0.044/\sqrt{E(MeV)}$ \cite{Moskal2014a}.


\section{Materials and methods}
\label{materialsandmethods}

The NEMA characteristics were estimated with the Geant4 Application for Tomographic Emission (GATE) simulation toolkit \cite{Jan2004, Jan2011}.
The GATE software is based on the Geant4 framework \cite{Agostinelli2003}, which is used to simulate interactions between radiation and matter.
The GATE software was used to simulate a~set of tomograph geometries with back-to-back two-gamma sources (defined by the points of the electron-positron annihilations), corresponding to sources defined in the NEMA norm.
Both primary and secondary scatterings of annihilation photons in the detector material were taken into account.
Simulated times and positions of interactions were smeared using tomograph resolutions obtained experimentally.
Further on, in order to reduce the fraction of events with gamma photons scattered in the phantom or in the detector, the signals were processed using selection criteria based on the correlations between the hit-time, hit-position and energy deposition of annihilation photons in the detector.

\subsection{Main parameters of the simulated tomograph}
\label{simulations}

In this article we study the NEMA characteristics of the J-PET tomograph as a~function of its length, diameter, number of detection layers, thickness of plastic strips and for three readouts: vacuum tube photomultipliers (PMT), silicon photomultipliers matrices (SiPM) and SiPMs combined with an additional wavelength shifters (WLS) layer \cite{Smyrski2017}.
Three diameters D~of the detector chamber (75~cm, 85~cm and 95~cm), three lengths L (20~cm, 50~cm and 100~cm) and two thicknesses T (4~mm and 7~mm) of scintillators are taken into account, for both single- or double-layer geometries.
Values of the diameter were chosen as typical for the presently available tomographs \cite{Slomka2016} and the axial field of view L~will be tested in the range from the typical present tomographs (about 20~cm) to 100~cm.
The diameter D (see Fig.~\ref{geometry_32strips}) is defined as a~distance between inner walls of the opposite strips.
The number of strips depends on the diameter of the scanner and is calculated as the number of edges of the regular polygon circumscribed around the ring with the radius D/2 (see Tab.~\ref{nr_of_strips}).


\begin{table}[!htb]
\begin{center}
\footnotesize

\begin{tabular}{|c|c|c|c|}
\hline
	\multirow{2}{*}{\textbf{Thickness T [mm]}} &
    \multicolumn{3}{c}{\textbf{Diameter D [cm]}} \vline \\\cline{2-4}
	&
	\textbf{75} &
	\textbf{85} &
	\textbf{95} \\
\hline
	\textbf{4} &
	590 &
	668 &
	746 \\
\hline
	\textbf{7} &
	336 &
	382 &
	426 \\
\hline
\end{tabular}

\caption{Number of strips in a~single layer of the detecting chamber as a~function of the thickness of the strip and the diameter of the chamber}
\label{nr_of_strips}
\end{center}
\end{table}

\begin{figure}[!htb]
\begin{center}
\includegraphics[width=0.49\textwidth]{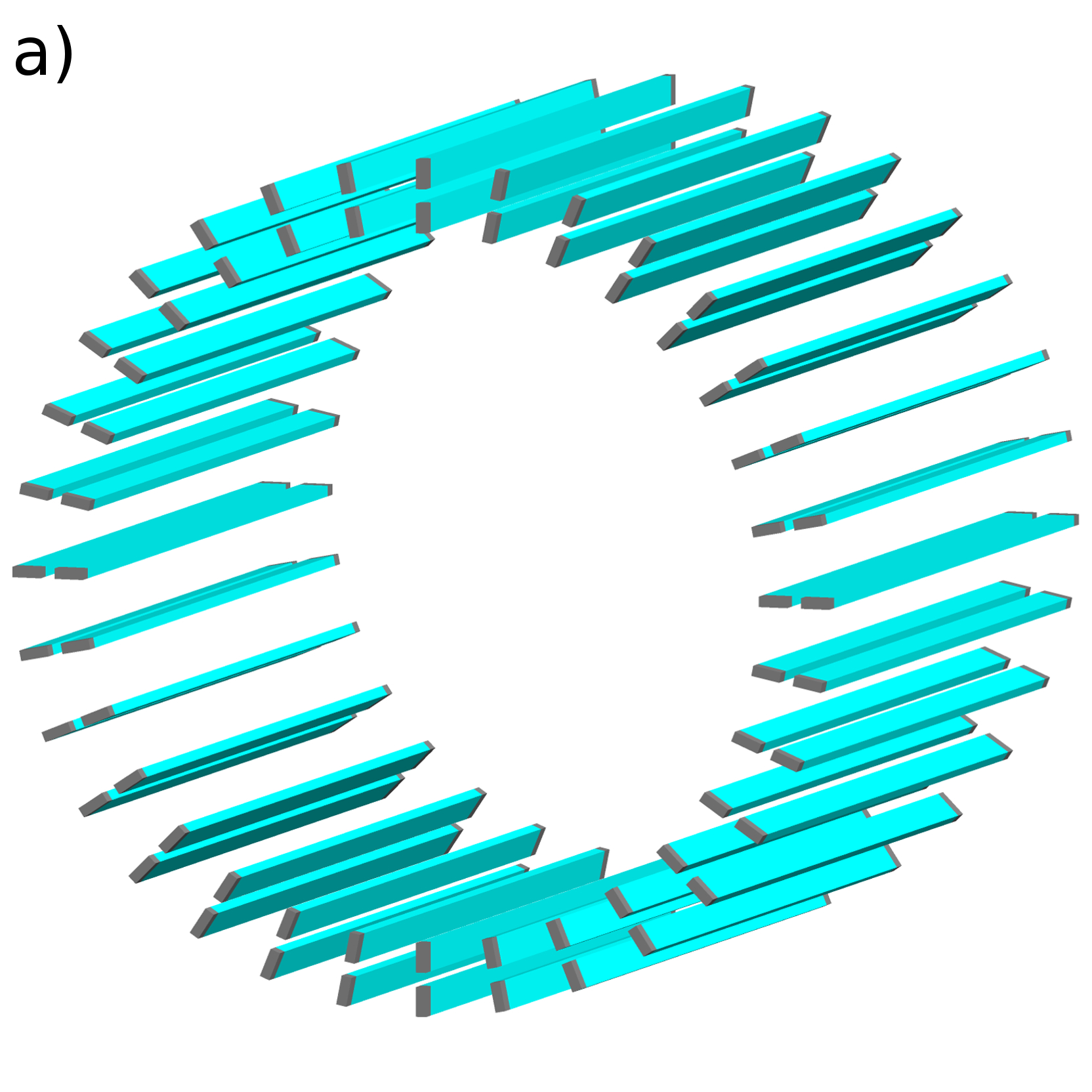}
\includegraphics[width=0.49\textwidth]{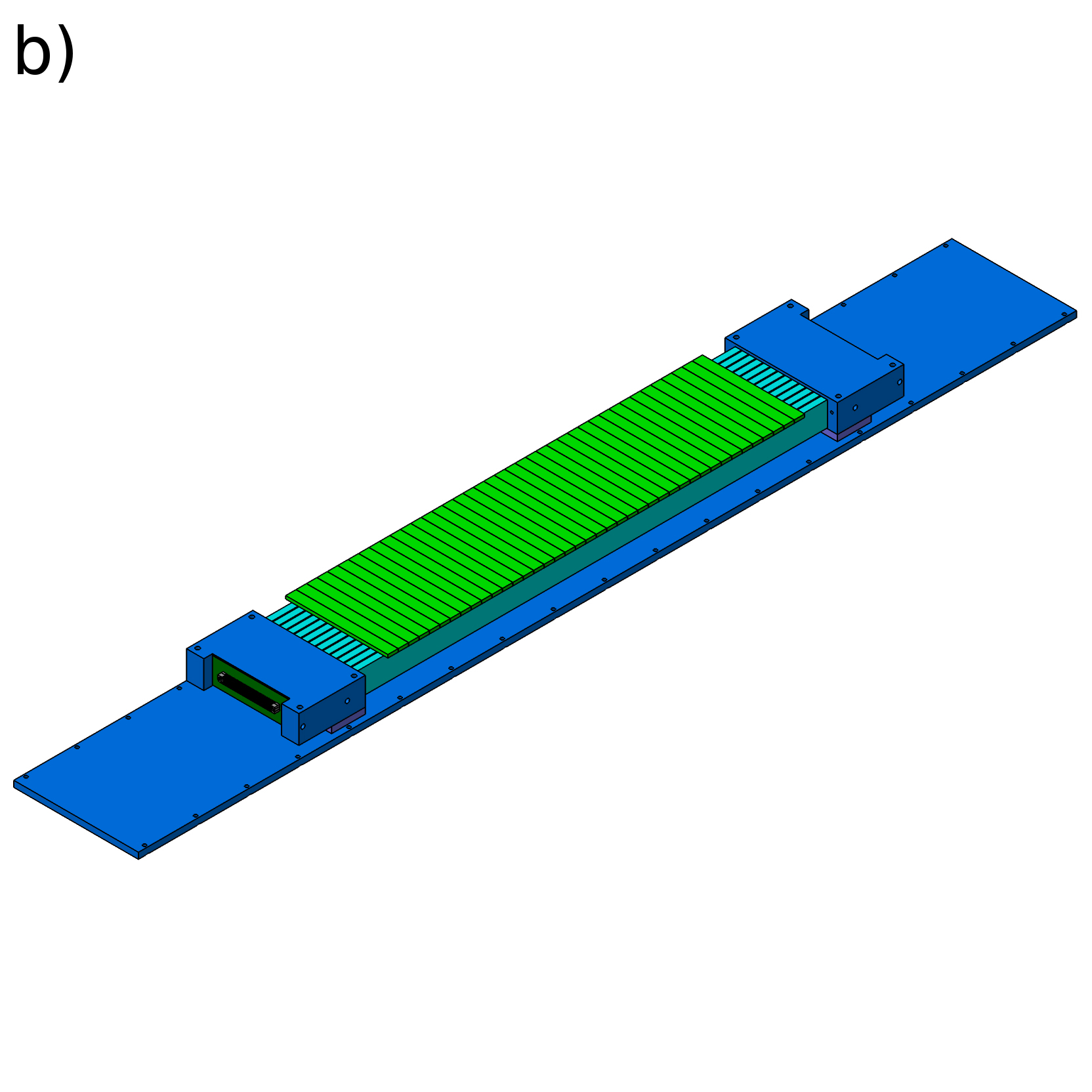}
\end{center}
\caption{
a)~The exemplary geometry of the 2-layer J-PET scanner.
For better visualization only 32~strips in each layer are shown.
Simulations of fully filled cylinder were performed.
b)~The geometry of the prototype module for the scanner based on both SiPM photomultipliers and WLS strips.
Scintillator strips (light blue) in the prototype module are arranged into clusters consisting of over a~dozen of strips.
The WLS strips (green) are arranged perpendicularly to the scintillator strips.
The results of the first experimental tests performed for the single plastic strip with array of WLS strips are reported in Ref.~\cite{Smyrski2017}.
}
\label{geometry_2layers_wls}
\end{figure}

For the 2-layer geometries, the second layer consists of the same number of strips arranged in the cylinder with the radius larger by 3~cm (Fig.~\ref{geometry_2layers_wls}a).
The diameter of double-layer geometry is defined as a~distance between inner walls of the opposite strips of the inner layer.

The depth (size of the strip along the radius of the scanner) of each scintillator is 20~mm and its thickness is 4~mm or 7~mm.
Presently used crystalline tomographs have crystals with depths from 20~mm up to 30~mm, while their cross-sections (perpendicular to the radius of the scanner) range from 4~x~4 mm$^2$ to 6.3 x 6.3 mm$^2$ \cite{Slomka2016, Vandenberghe2016}.
The coincidence resolving time (CRT) and axial spatial resolution was simulated for three different readout solutions, referred to as the vacuum tube photomultipliers, silicon photomultipliers matrices or the SiPM readout with additional layer of the WLS strips (Fig.~\ref{geometry_2layers_wls}b).
The WLS strips were arranged perpendicularly to the scintillator strips allowing for the determination of the gamma photon interaction point along the tomograph axis based on the distribution of amplitudes of light signals in WLS strips~\cite{Smyrski2017}.
An analogical solution was proposed for the AX-PET detector~\cite{Casella2014, Gillam2014, Solevi2015}.

\begin{figure}[!htb]
\begin{center}
\includegraphics[width=0.49\textwidth]{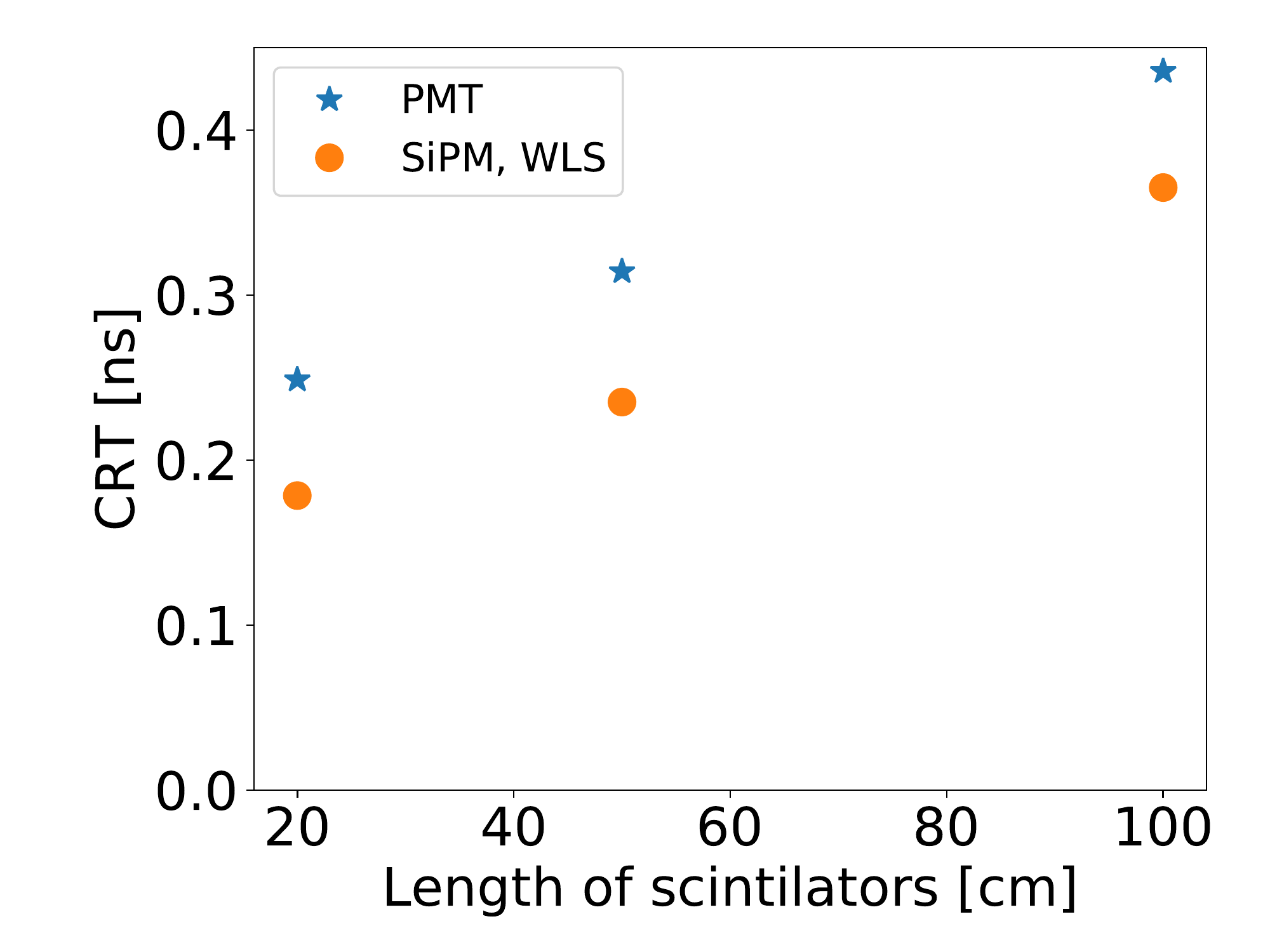}
\includegraphics[width=0.49\textwidth]{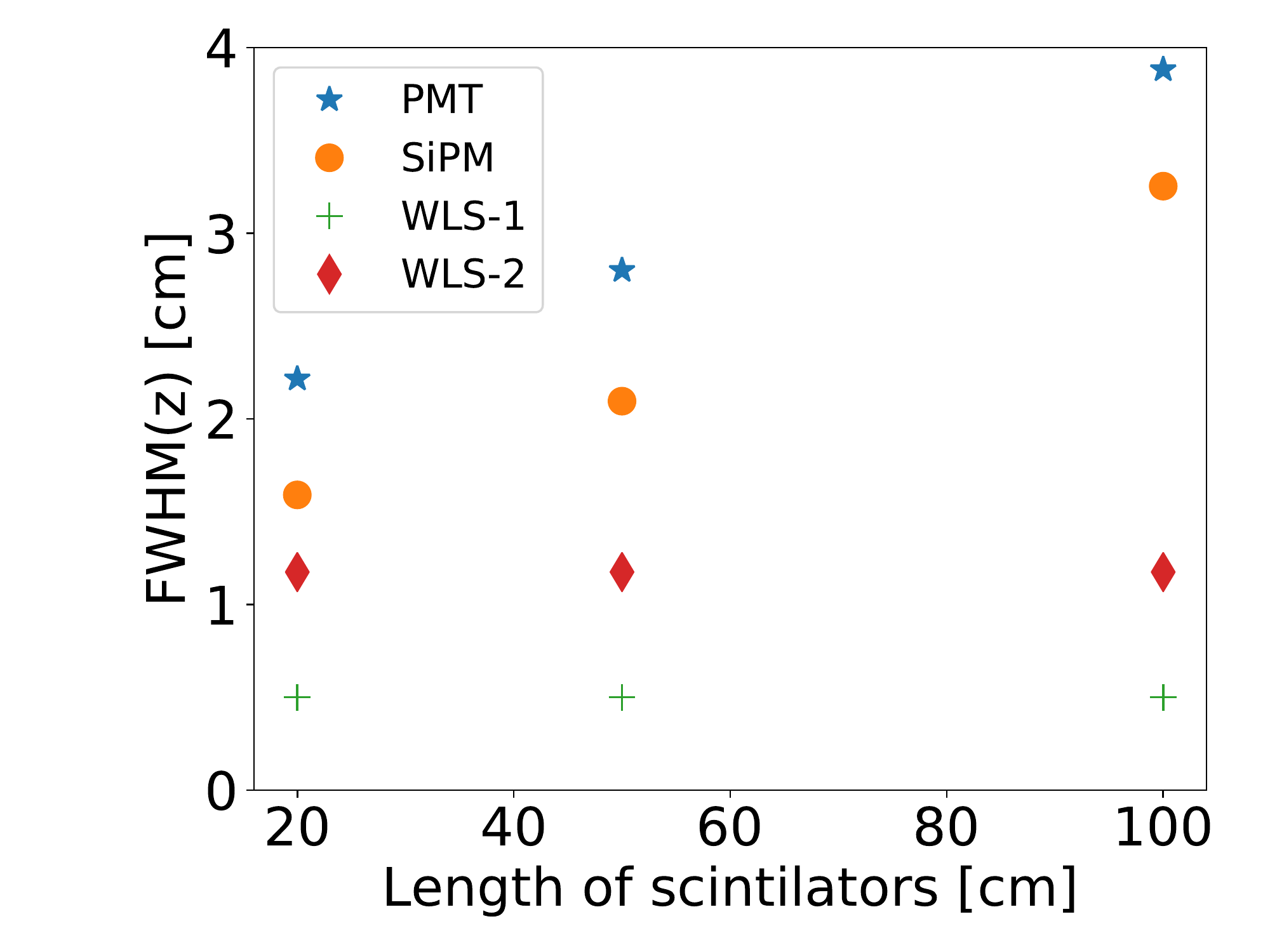}
\end{center}
\caption{
The values of CRT (left) and axial resolution (right), used for the simulations, as a~function of the length of the scintillator, shown for different readouts \cite{Moskal2016, Smyrski2017}.
More details are in the text.
}
\label{crt}
\end{figure}

Configurations of all simulated cases are listed in Tab.~\ref{possible_configurations} and the used values of CRT and axial resolution as a~function of the tomograph length L~are shown in Fig.~\ref{crt}.
The values of CRT and FWHM (in z coordinate) for PMTs and SiPMs were estimated based on simulations presented in Ref.~\cite{Moskal2016} tuned to the empirical results for single- and double-strip J-PET prototypes \cite{Moskal2014a, Moskal2015}.
The FWHM with the additional layer of the WLS strips was measured with the test setup described recently in the article \cite{Smyrski2017}.
The values indicated as WLS-2 show resolution achieved in the first test \cite{Smyrski2017}.
However, since the system was not fully optimized, there is still room for significant improvement which is indicated in Fig.~\ref{crt} as WLS-1.
The improvement may be achieved by better matching between the emission spectrum of the scintillator and the absorption spectrum of the WLS and more efficient photon readout.
In the test, only half of the WLS surface was covered by the SiPM.

Simulations were performed using the list of physics processes called \textit{emlivermore\_polar} \cite{Geant4}.
This model is designed for any applications required higher accuracy of electrons, hadrons and ion tracking without magnetic field.
Physical processes included in simulations are the photoelectric effect, Compton scattering, gamma conversion, Rayleigh scattering, ionisation and bremsstrahlung.
Polarization of gamma photons is also simulated.
The model describes the interactions of electrons and photons with matter down to 10~eV and up to 100~GeV using interpolated data tables based on the Livermore library \cite{Geant4}.
This range of energies is well chosen for the purpose of the J-PET simulations due to the low energy cut on 10~keV.
Energy thresholds and other selection criteria used in analysis are described in the following section.

\begin{table}
\begin{center}

\begingroup
\renewcommand\arraystretch{1.3}
\noindent\begin{tabular}[t]{
   >{\centering\arraybackslash}m{2cm}
   >{\centering\arraybackslash}m{0.5cm}
   >{\centering\arraybackslash}m{2cm}
   >{\centering\arraybackslash}m{0.5cm}
   >{\centering\arraybackslash}m{2cm}
   >{\centering\arraybackslash}m{0.5cm}
   >{\centering\arraybackslash}m{2cm}
   >{\centering\arraybackslash}m{0.5cm}
   >{\centering\arraybackslash}m{2cm}
}
  \begin{tabular}[t]{|c|}
	\hline	
	Layers \\
	\hline
	1 \\
	2 \\
	\hline
  \end{tabular}
&
	x 
&  
  \begin{tabular}[t]{|c|}
	\hline	
	Thickness \\
	\hline
	4 mm \\
	7 mm \\
	\hline
  \end{tabular}
&
	x 
&  
  \begin{tabular}[t]{|c|}
	\hline	
	L [cm] \\
	\hline
	20 \\
	50 \\
	100 \\
	\hline
  \end{tabular}
&
	x
&  
  \begin{tabular}[t]{|c|}
	\hline	
	D [cm]\\
	\hline
	75 \\
	85 \\
	95 \\
	\hline
  \end{tabular}
&
	x 
&  
  \begin{tabular}[t]{|c|}
	\hline	
	Readout \\
	\hline
	PMT \\
	SiPM \\
	SiPM+WLS \\
	\hline
  \end{tabular}
\end{tabular}
\endgroup

\caption{Configurations of simulated detecting systems which may differ with number of layers of the detector and their diameters, thickness, length of the scintillator strip and type of readout.}
\label{possible_configurations}
\end{center}
\end{table}


\FloatBarrier

\subsection{Event selection method}
\label{eventselection}

An event is defined as a~set of consecutive interactions of photons, originating from a~single $e^+e^-$ annihilation and all interactions of secondary particles.
In case of the \mbox{J-PET} detector, most of interactions are Compton scatterings.
If the interactions are detected within the fixed time coincidence window of 3 ns, they are operationally considered to originate from the same annihilation event.
Any two scatterings within an event may form a~coincidence. Coincidences may be classified into three types: true, scattered and accidental.

\begin{figure}[!htb]
\begin{center}
\includegraphics[width=0.33\textwidth]{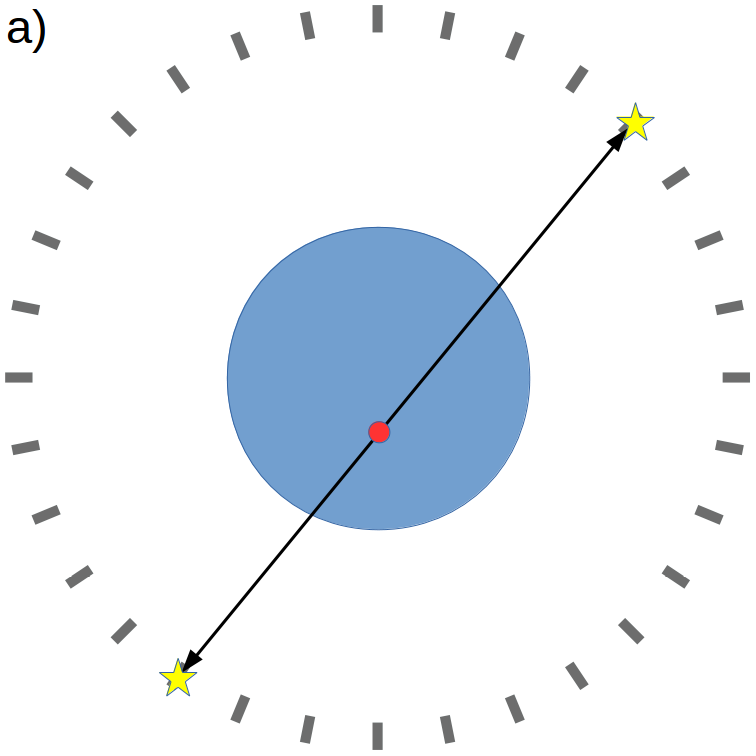}
\includegraphics[width=0.33\textwidth]{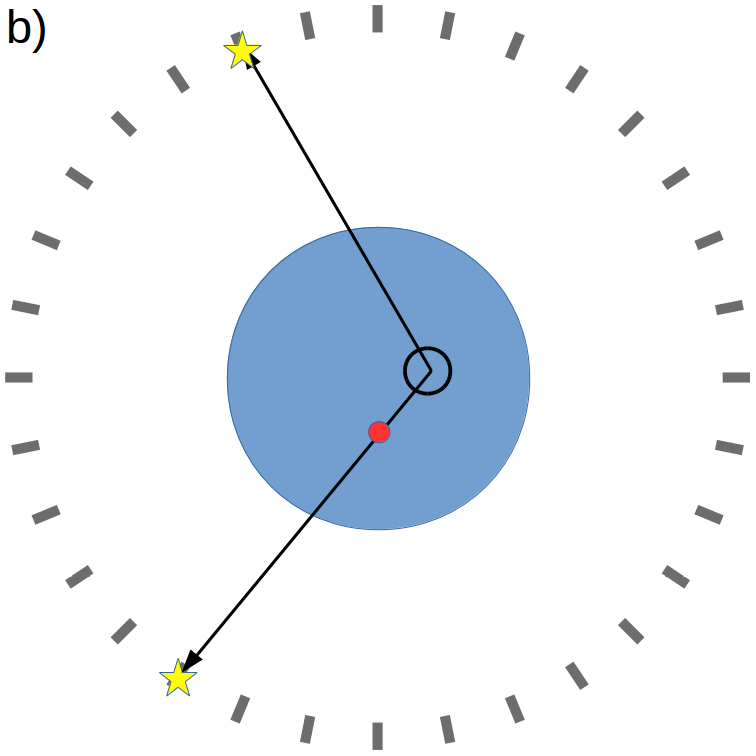}
\includegraphics[width=0.33\textwidth]{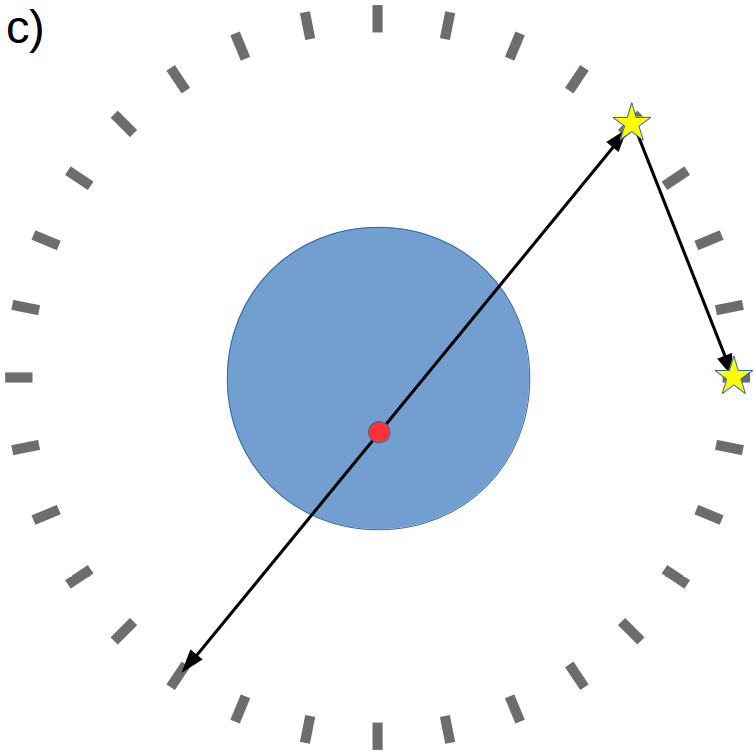}
\includegraphics[width=0.33\textwidth]{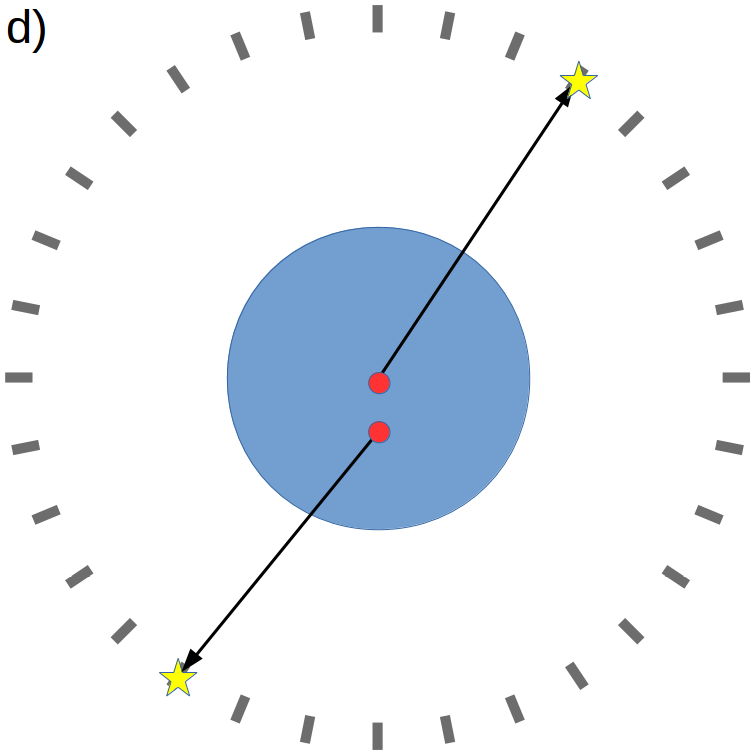}
\end{center}
\caption{
Pictorial definitions of different types of coincidences: a) true coincidence, b) phantom-scattered coincidence, c) detector-scattered coincidence and d) accidental coincidence.
Blue large circle denotes the phantom, red dots denote the annihilation points, black circle denotes the scattering place in the phantom and yellow stars indicate interactions of gamma photons in the detector.
}
\label{coincidences1}
\end{figure}

While true coincidences are desirable, the scattered and accidental coincidences contribute to the background, hinder the reconstruction of the image and decrease its final quality.
Scattered coincidences may be divided into the detector- and phantom-scattered coincidences.
Detector-scattered coincidences are defined as coincidences in which at least one of the interactions does not originate from the annihilation photon but from another (often primary) scattering in the detector.
Phantom-scattered coincidences are those in which at least one of the annihilation photons was scattered in the phantom before the detection.
Coincidences of scatterings of photons from different annihilations are called accidental or random.
Pictorial definitions of these different types of coincidences are shown in Fig. \ref{coincidences1}.

\begin{figure}[!htb]
\begin{center}
\includegraphics[width=0.32\textwidth]{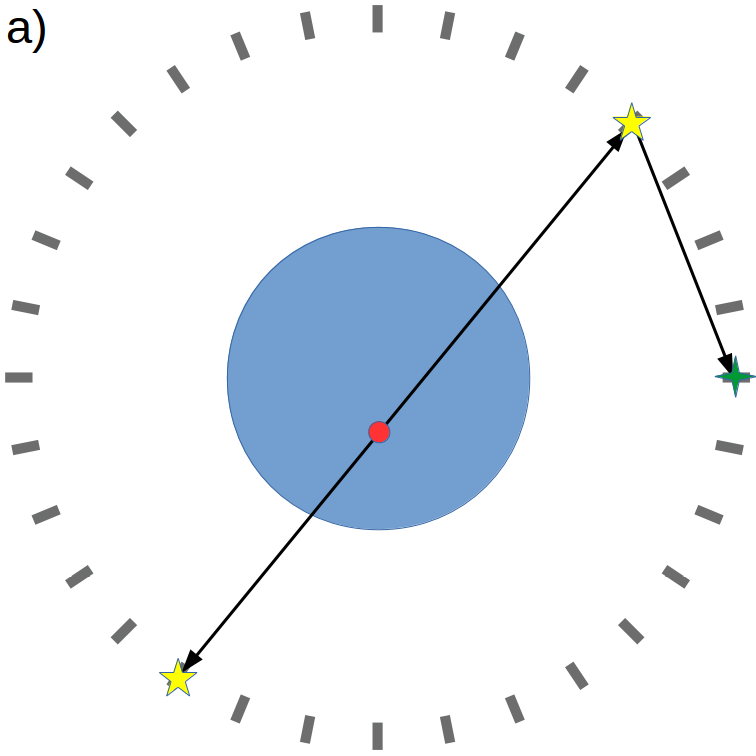}
\includegraphics[width=0.32\textwidth]{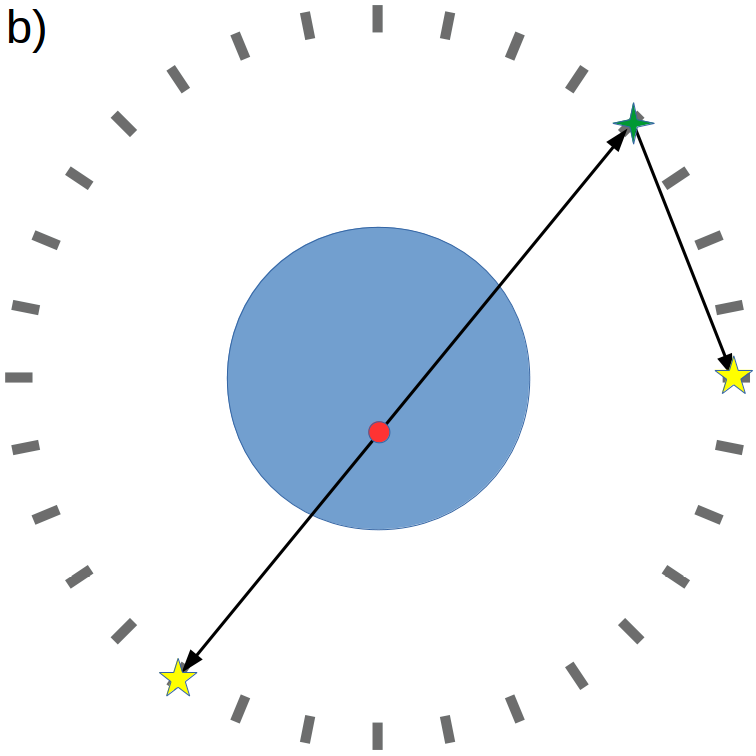}
\includegraphics[width=0.32\textwidth]{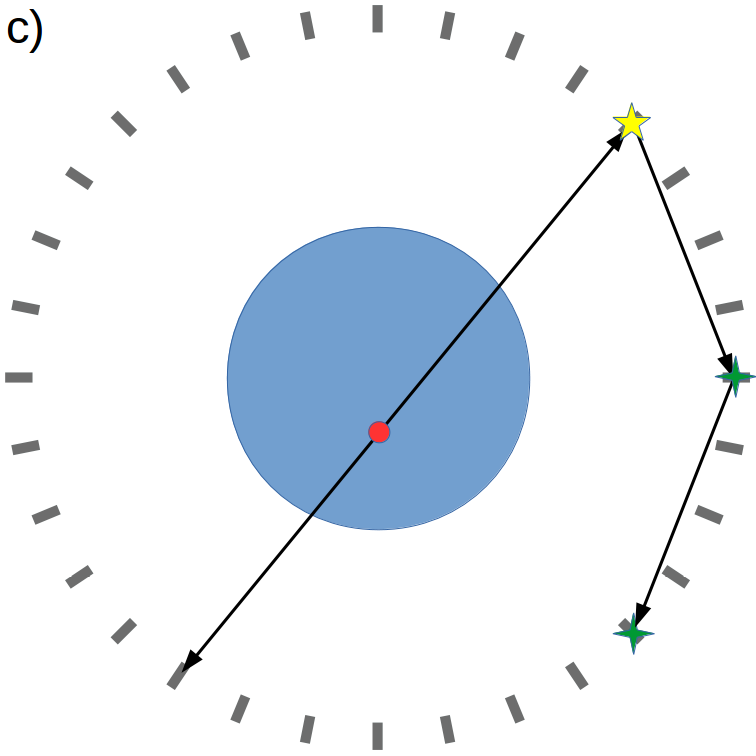}
\end{center}
\caption{
Exemplary events with three depositions of energy above the noise threshold: a) true coincidence, b) and c) detector-scattered coincidences.
Yellow 5-arm stars visualize depositions of energy bigger than the fixed energy threshold (200~keV) and the green 4-arm stars visualize depositions of energy lower than 200~keV but bigger than the noise energy threshold equal to 10~keV.
}
\label{coincidences2}
\end{figure}

In order to extract true coincidences from the set of all coincidences, a~two-level selection procedure is performed.
In the first step, only events with exactly two interactions registered with energy loss larger than 200~keV each and any number of interactions with energy loss larger than 10~keV (above the noise level) and smaller than 200~keV are accepted.
We accept events with more than two scatterings in order to avoid rejection of such events as shown in Fig.~\ref{coincidences2}a, which can be classified as true coincidences.
However, these conditions lead to the acceptance of unwanted coincidences shown e.g. in Fig.~\ref{coincidences2}b and \ref{coincidences2}c.
Yet, the requirement of at least two interactions with energy loss above the 200~keV threshold reduces to the negligible level coincidences of the type shown in Fig.~\ref{coincidences2}c.
This is because the 511~keV gamma photon cannot deposit more than 184~keV in more than one scattering \cite{Kowalski2016}.

\begin{figure}[!htb]
\begin{center}
\includegraphics[width=0.5\textwidth]{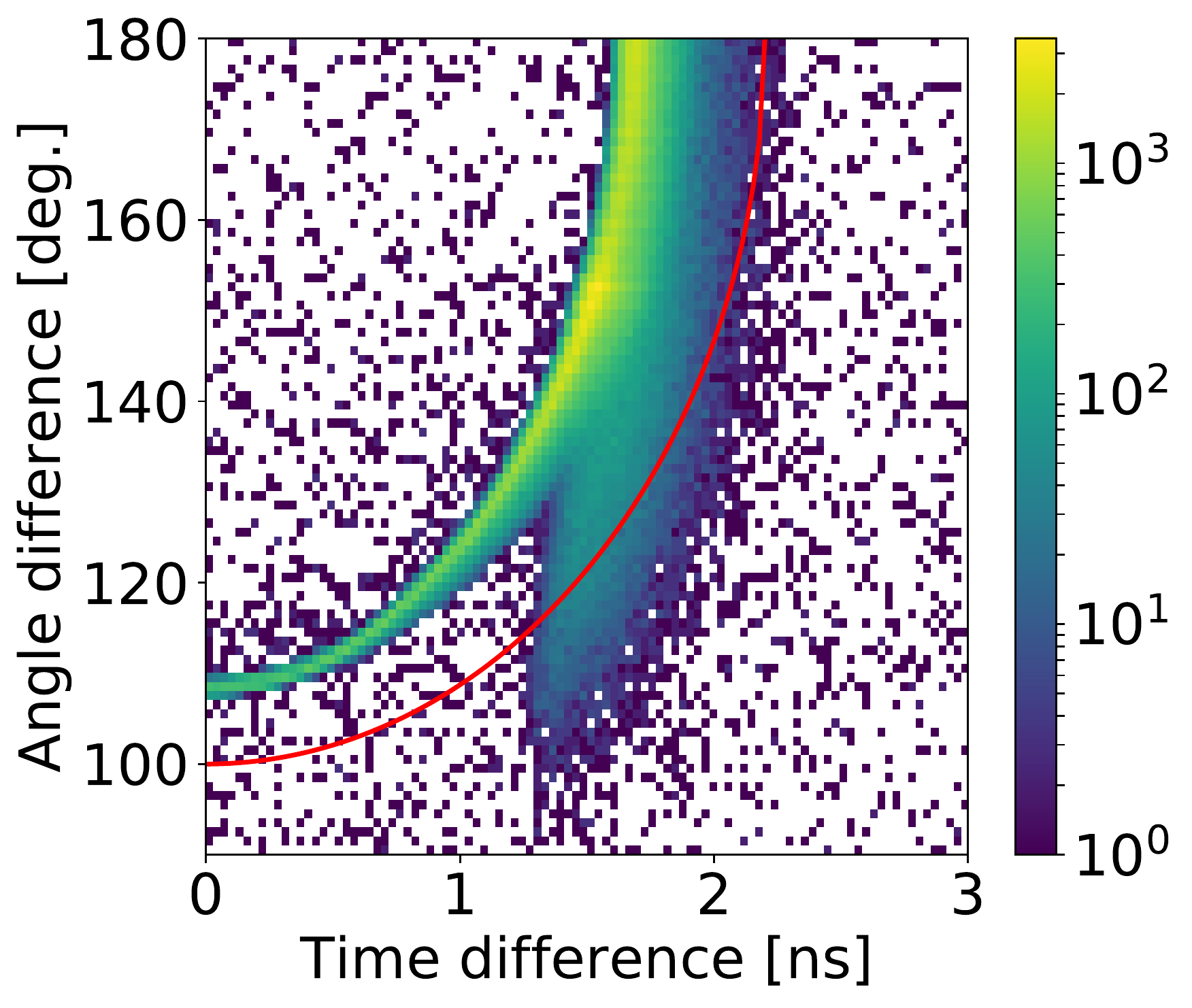}
\includegraphics[width=0.44\textwidth]{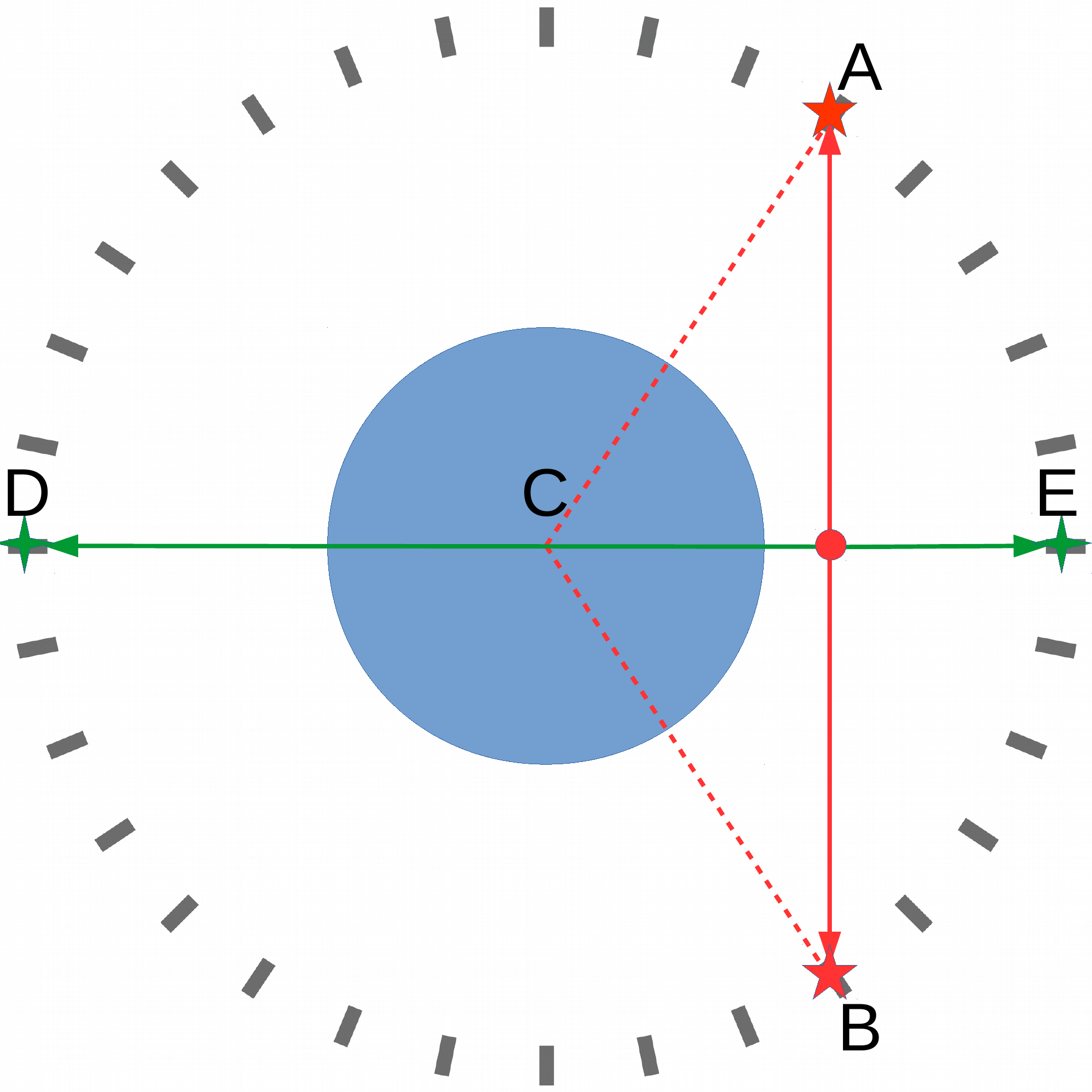}
\end{center}
\caption{
(left) Scatter plot of angle differences vs. time differences for annihilations from the 70~cm long rod source placed axially 25~cm from the tomograph axis.
In the centre of the scanner, cylindrical phantom was placed (more details in text).
(right) Schematic cross-section of one layer of the tomograph with cylindrical phantom (large blue circle) and the source (red dot).
Two extreme cases explaining the main structure seen in (left) panel are indicated.
For the LOR passing through the centre of the tomograph (green arrows), the relative angle between strips is equal to 180$^{\circ}$ (angle DCE) and the hit time difference is equal to ${50~cm \over c} \approx 1.5~ns$.
In the other extreme case (red arrows) the time difference between hits is equal to zero (paths of two gamma photons are equal) and the difference of angles between two red dotted lines is equal to about 115$^{\circ}$ (angle ACB), as illustrated in the figure.
}
\label{secondlevelselection}
\end{figure}

At the second level of the event selection, information about times and azimuthal angles\footnote{
The azimuthal angle is the central angle in the plane perpendicular to the axis of the scanner.
In Fig.~\ref{secondlevelselection} the azimuthal angle of the scintillator A~would be ACE.
} of the scintillator strips is used.
Times and azimuthal angles differences are calculated for each pair of hits.
Using calculated values, part of events is rejected.
The criterion of selection is based on the simulation with the cylindrical phantom (NEMA scatter phantom) and the linear source placed at the radial distance of 25~cm from the axis of the scanner \cite{Kowalski2016}.
The source was a~70~cm long rod with diameter of 3.2~mm and activity of 1~MBq.
The simulated geometry consisted of a~single scintillator layer with 50~cm long and 7~mm thick strips and the diameter of 85~cm.
The simulation showed that all true coincidences lie in a~well-defined region in the space spanned by the times and azimuthal angles differences.
This region, which lies above the red line\footnote{
The red line is the ellipse with vertex at (2.2 ns, 180$^{\circ}$), co-vertex at (0 ns, 80$^{\circ}$) and center at (0 ns, 180$^{\circ}$).
The choice is based on results of simulations shown in Fig.~\ref{secondlevelselection}.
The choice of the red-line criterion is rather conservative and leaves room for further reduction of scatter fraction in case of smaller objects imaging.
} in Fig.~\ref{secondlevelselection}, may be defined as a~selection criterion.
The longitudinal structure visible for time difference near to 1.5~ns is caused by the phantom-scattered coincidences and it was discussed in Ref. \cite{Kowalski2015}.

\begin{figure}[!htb]
\begin{center}
\includegraphics[width=200pt]{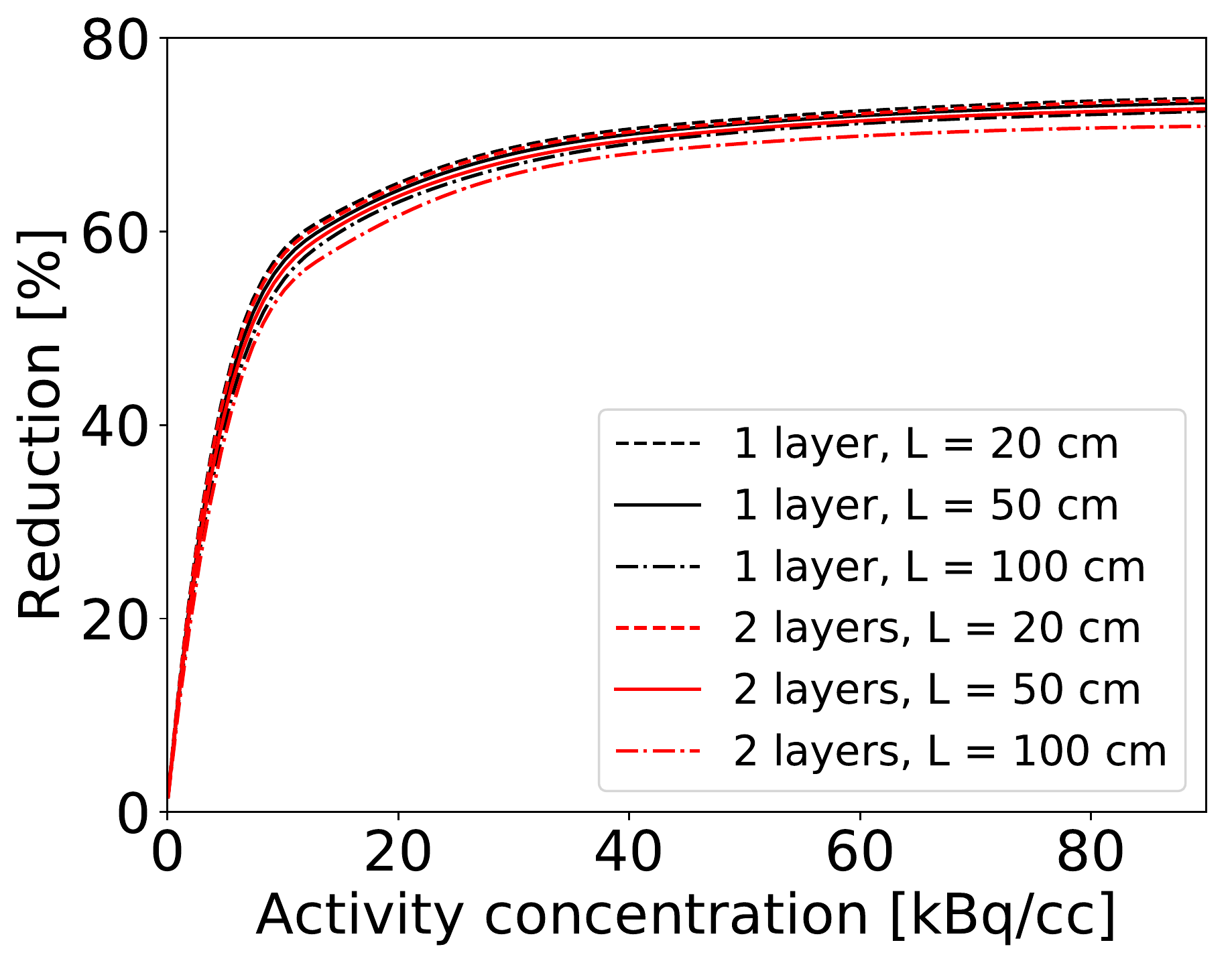}
\end{center}
\caption{
Reduction of number of events provided by the second level of event selection method for different geometries of the J-PET scanner
}
\label{second_lvl_selection_reduction}
\end{figure}

\begin{figure}[!htb]
\begin{center}
\includegraphics[width=0.49\textwidth]{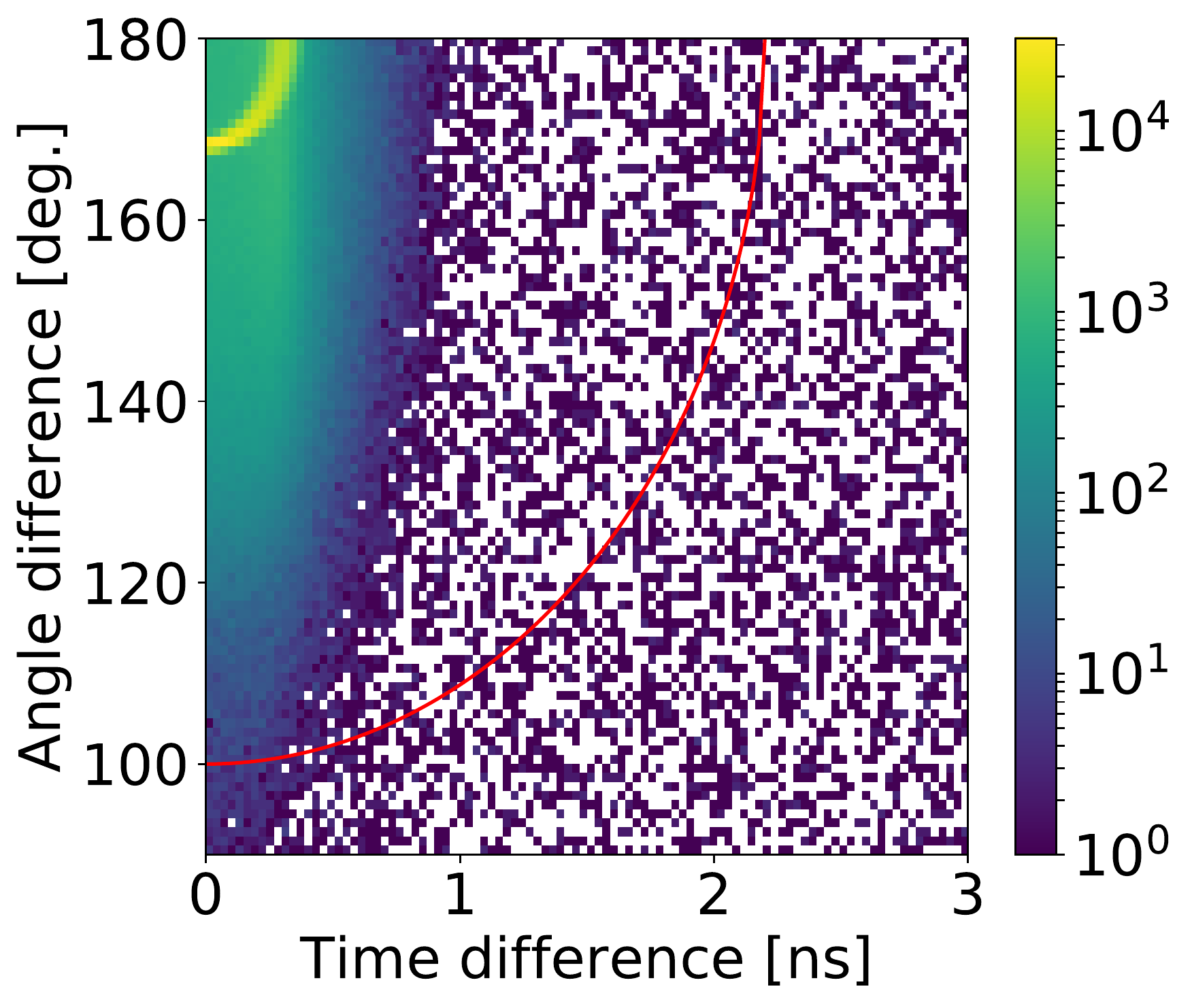}
\includegraphics[width=0.49\textwidth]{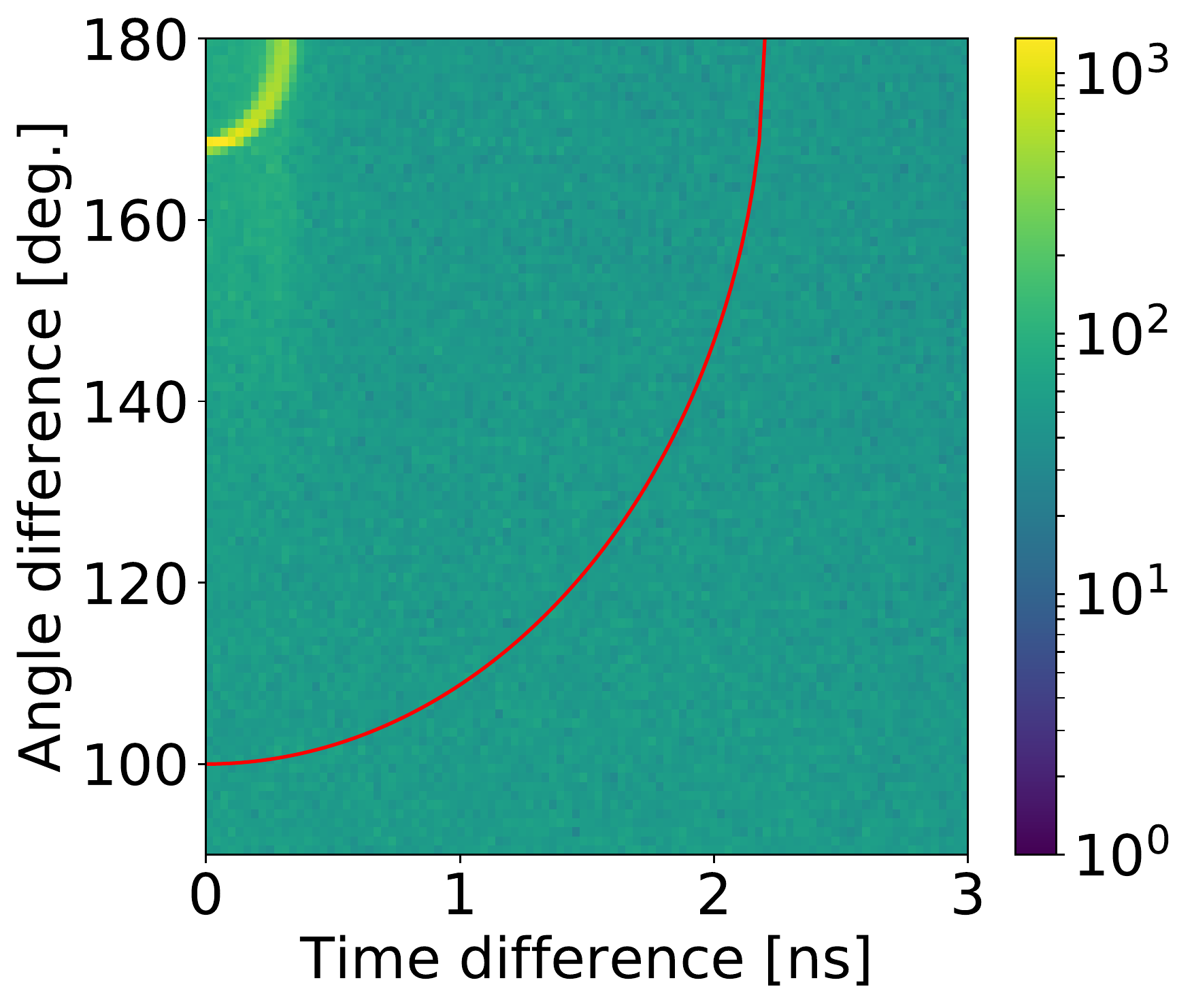}
\end{center}
\caption{
Results of the NECR characteristic simulation for the geometry with 50~cm length scintillating strips visualized in the form of scatter plots of angle differences vs. time differences.
The source was the 70~cm long rod placed axially inside the scatter phantom at distance of 4.5~cm from the axis of the tomograph.
Left panel shows result of simulation for small activity of 45 Bq/cc  and right panel for high activity of 90.9 kBq/cc.
The second level selection criterion indicated as the red line cause reduction of events by a~factor of 0.5\% to 70\% for the left and right panel, respectively.
}
\label{second_lvl_selection_examples}
\end{figure}

The selection criteria discussed above allow to filter out the most of scattered coincidences from true coincidences.
The method is based on the values of deposited energies and times and azimuthal angles differences.
Number of reduced events in the second level selection depends on the simulation type and the source activity.
For the simulation with rod source placed at 25 cm from the axis of the scanner (Fig. \ref{secondlevelselection}), the percentage of rejected events (reduction factor with respect to the first level of selection) was 2.3\%.
In case of spatial resolution simulations (section \ref{sr_section}) the reduction was 0.5\%.
For the NECR simulations (section \ref{sf_section}), the reduction was between 0.5\% for small activity concentrations and about 70\% for highest activity concentrations (about 90 kBq/cc), which shows that the higher the activity, the more important the reduction provided by the second level of event selection method (Figs\ref{second_lvl_selection_reduction}-\ref{second_lvl_selection_examples}).
The chosen selection criterion may be also adjusted to the size of the investigated object and to the geometrical properties of the scanner.

\FloatBarrier

\subsection{Sensitivity}
\label{materials_sensitivity}

The sensitivity of a~positron emission tomograph is expressed as the true coincidence events rate T~normalized to the total activity A~of the source.
The selection criteria for the true coincidence were described in section \ref{eventselection}.
In particular we require that each annihilation photon deposits at least 200 keV.
This criterion limits registration of photons scattered in the patient to the range from 0 to 60 degrees \cite{Moskal2016}.
In order to calculate the sensitivity, a~linear 1~MBq source of back-to-back gamma photons with length of 70~cm was simulated along the axis of the scanner in the centre of the AFOV.
The NEMA norm requires that the activity should be such, that the number of accidental coincidences is smaller than 5\% of all prompt coincidences.
The activity of 1~MBq fulfills that condition for all simulated geometries.
The ratio of the accidental coincidences ranges from 0.80\% to 1.34\%, depending on the geometry.

\FloatBarrier

\subsection{Spatial resolution}
\label{materials_resolution}

The spatial resolution of a~PET scanner represents its ability to distinguish between two points after image reconstruction \cite{NEMA}.
In order to obtain this characteristic, signal determination from the annihilation must be performed and corresponding images must be reconstructed.
The Full Width at Half Maximum (FWHM) of the obtained distributions is referred to as Point Spread Function (PSF) and is used as a~measure of the spatial resolution.

Since the spatial resolution depends on the position inside the AFOV of the scanner, the PSF must be determined for six different, defined by the norm, positions.
In the axial direction, source should be placed at the centre of the AFOV and at the distance of three-eighths of the length of the AFOV from the centre of the scanner.
In the transverse (i.e. radial) direction, source should be placed at distances: 1~cm, 10~cm and 20~cm.

In simulations, we used a~back-to-back gamma source with a~diameter of 1~mm.
The activity of the source was sufficiently low (equal to 370 kBq [10 $\mu$Ci]) in order to fulfill the condition that the number of accidental coincidences should be less than 5\% of all collected events.
The ratio of the accidental coincidences for each simulated geometry and source position was smaller than 0.2\%.
According to the NEMA norm, the number of collected prompt coincidences for each position of the source should be at least 100,000.

After collecting required number of events, a~set of coincidence events was selected using the method described in section \ref{eventselection}.
Before reconstruction, the simulated data were smeared taking into account experimental time and position resolution.
After the selection and data smearing, reconstruction was performed.
The smearing of position and time was performed for three different readout cases as it is shown in Fig.~\ref{crt}.
For the reconstruction, according to the NEMA norm, the Filtered Back-Projection (FBP) method was used.
The 3D~FBP algorithm from the Software for Tomographic Image Reconstruction (STIR) \cite{Thielemans2012} package was used \cite{Shopa2017}.

For each 3D~image, a~voxel with the maximum intensity was found and three \mbox{1-dimensional} profiles going through this voxel in each directions (x, y, z) were determined.
For each profile, the values of FWHM were obtained.
These values are interpreted as the spatial resolution of the scanner.
Because of the variety of possible configurations of the J-PET scanner (taking into account parameters listed in section~\ref{simulations}) only exemplary results for the single-layer geometries are presented in the article.
Since the additional layer of strips influences mainly the detection efficiency, similar results should be expected for the double-layer geometries.

\subsection{Scatter fraction}
\label{materials_sf}

The scatter fraction of the PET scanner quantifies the sensitivity of the detector to scattered radiation.
It may be simply expressed as a~ratio between the scattered coincidences and the sum of the scattered and the true coincidences.
In case of the J-PET detector, built from plastic scintillators, the annihilation photon interacts via Compton scattering and there is a~significant fraction of events (depending on the detector geometry) when the annihilation photon undergoes more than one scattering in the detector.
Therefore, in case of the J-PET tomograph, the scattered coincidences consist of the detector- and phantom-scattered coincidences.
However, applying the fixed energy threshold (200~keV) causes that most of the detector-scattered coincidences is reduced (see section~\ref{eventselection} and Ref.~\cite{Kowalski2016} for more details).

The simulated phantom is a~solid cylinder made of polyethylene with an outside diameter equal to 203~mm and length 700~mm.
Parallel to the axis of the cylinder, a~hole with diameter 6.4~mm is drilled in a~radial distance from the axis of the phantom equal to 45~mm.
A~line source insert is also made of polyethylene and it is a~tube with the inside diameter 3.2~mm and outside diameter 4.8~mm.
The tube may represent known activity and be placed inside the hole of the phantom.

In the simulation used to obtain the scatter fraction of the J-PET scanner, the source of back-to-back annihilation photons with activity 1~kBq was generated.
The value of the activity for obtaining the scatter fraction is limited by the condition that the ratio between the random and true coincidences should be smaller than 1\% \cite{NEMA}.
For such a~small activity, there were only single accidental coincidences per each data set consisting from hundreds of thousands of prompt coincidences, which means that the activity of 1~kBq fulfills this condition.
The NEMA norm also requires that the number of acquired prompt coincidences must be at least 500,000.
In our studies, the number of prompt coincidences was 1~mln.

\begin{figure}[!htb]
\begin{center}
\includegraphics[width=0.49\textwidth]{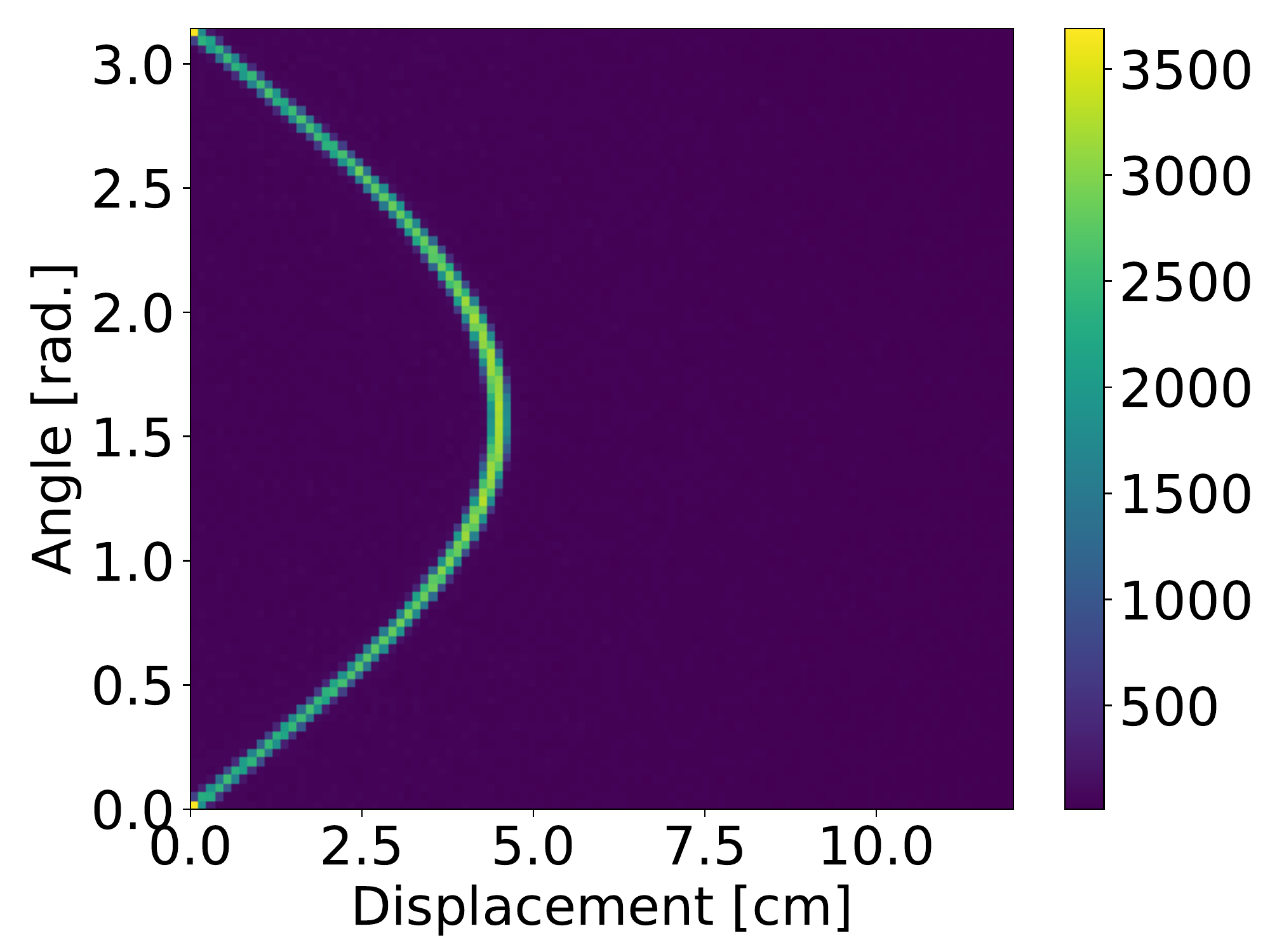}
\includegraphics[width=0.49\textwidth]{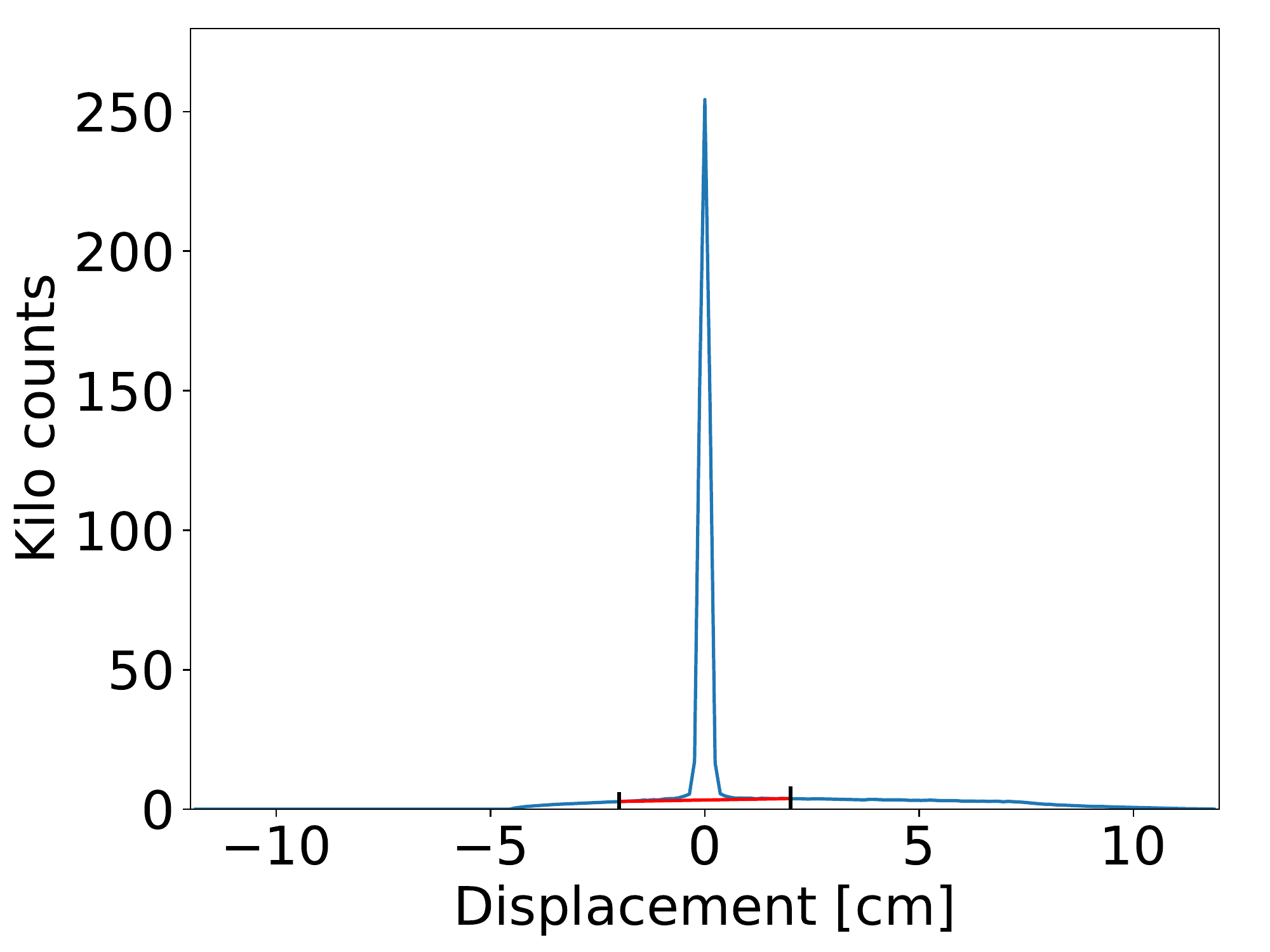}
\end{center}
\caption{
Results of the scatter fraction simulation for single-layer geometry with diameter of 85~cm and strips with length of 50~cm.
(left) The sinogram for a~whole scanner (all slices summed).
(right) One dimensional profile calculated as an aligned to zero using maximum value and summed projections of sinogram presented in left panel.
The cut level visualized as a~red line is determined using values of the profile at $\pm 2$ cm from zero.
}
\label{sinogramimage}
\end{figure}

As a~first step of the data processing, the space inside the scanner is axially divided into N~virtual slices and N$^2$ oblique sinograms are generated.
The sinogram is a~transformation of the line of response (LOR) into a~pair of values: the distance of this LOR to the centre of the detector (in x-y plane) and the angle between the LOR and the \textit{x}~axis of the cross-section of the scanner.
In the second stage, oblique sinograms are converted into rebinned sinograms using the Single Slice Rebinning (SSRB) algorithm.
The number of rebinned sinograms is equal to 2N-1.
After that, rebinned sinograms are merged into one sinogram (example in the left panel of Fig.~\ref{sinogramimage}).
Using this summed sinogram, all projections are aligned with maximum value to zero and summed in order to get a~one dimensional profile (example of such a~profile is shown in the right panel of Fig.~\ref{sinogramimage}).
After summing up, the values of such obtained profile at distances $\pm 2$~cm from zero are calculated.
The area of the profile over the line crossing two points at $\pm 2$~cm is treated as true coincidences, whereas the area below this line corresponds to the scattered (and accidental) coincidences.

The scatter fraction was calculated for 6~geometries: one or two scintillator layers and three lengths L~of scintillators: 20~cm, 50~cm and 100~cm.
The diameter of the detector chamber was 85~cm and the profile of the single scintillator was 20~mm~x~7~mm.

\subsection{Noise equivalent count rate}
\label{materials_necr}

The NECR is characteristic that shows the effect of a~correction for scattered and random coincidences as a~function of the source activity, and it is a~measure of the effective sensitivity of the scanner~\cite{Conti2009}.
The NECR may be defined as: $NECR = {T^2 \over {T+S+R}}$, where T~stands for the rate of true coincidences, S - scattered coincidences, R - random (accidental) coincidences.

The NECR characteristic is a~figure of merit showing the optimal value of activity of the source for a~fixed geometry of the scanner.
The optimal value is defined by the position of the NECR peak.
The smaller value of the activity at the peak, the smaller activity may be applied to the patient in order to get the best possible results.
The NECR is also related to the image quality SNR (Signal to Noise Ratio) \cite{Yang2015}.

The method of obtaining the NECR is similar to the method of measuring and calculating the scatter fraction.
While the scatter fraction is measured for single activity of the source, the NECR varies as a~function of the activity.


\section{Results}
\label{section_reults}

\FloatBarrier

\subsection{Sensitivity}
\label{sensitivity_section}

\begin{figure}[!htb]
\begin{center}
\includegraphics[width=0.495\textwidth]{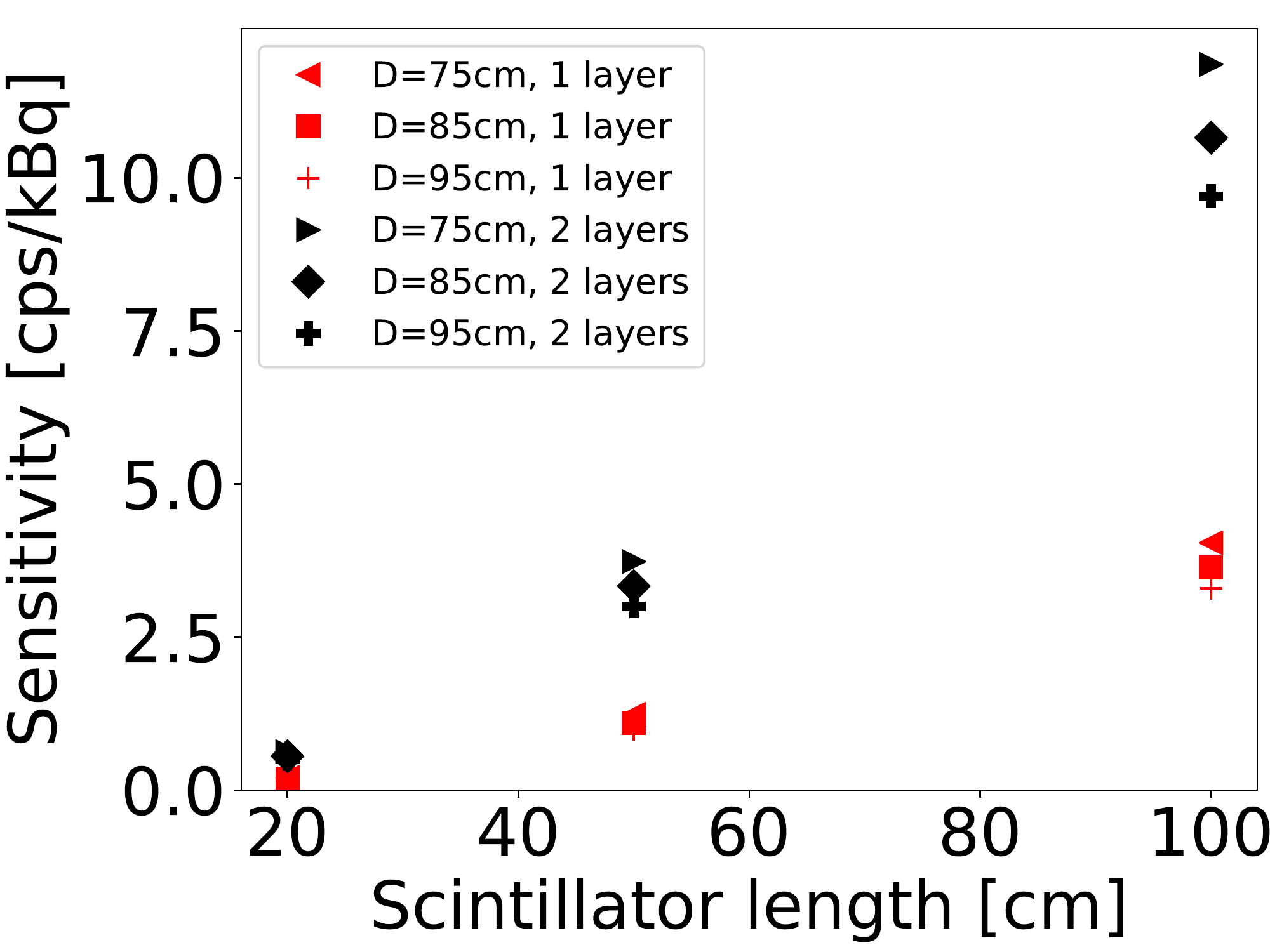}
\end{center}
\caption{
General sensitivities for different geometries of the J-PET scanner.
}
\label{sensitivity}
\end{figure}

The values of sensitivity for different geometries of the J-PET scanner are presented in Fig.~\ref{sensitivity}.
All events were grouped into slices in order to obtain sensitivity profiles.
Grouping was done using information from the simulation about exact position of place of emission of annihilation photons for each coincidence.

\begin{figure}[!htb]
\begin{center}

\includegraphics[width=0.325\textwidth]{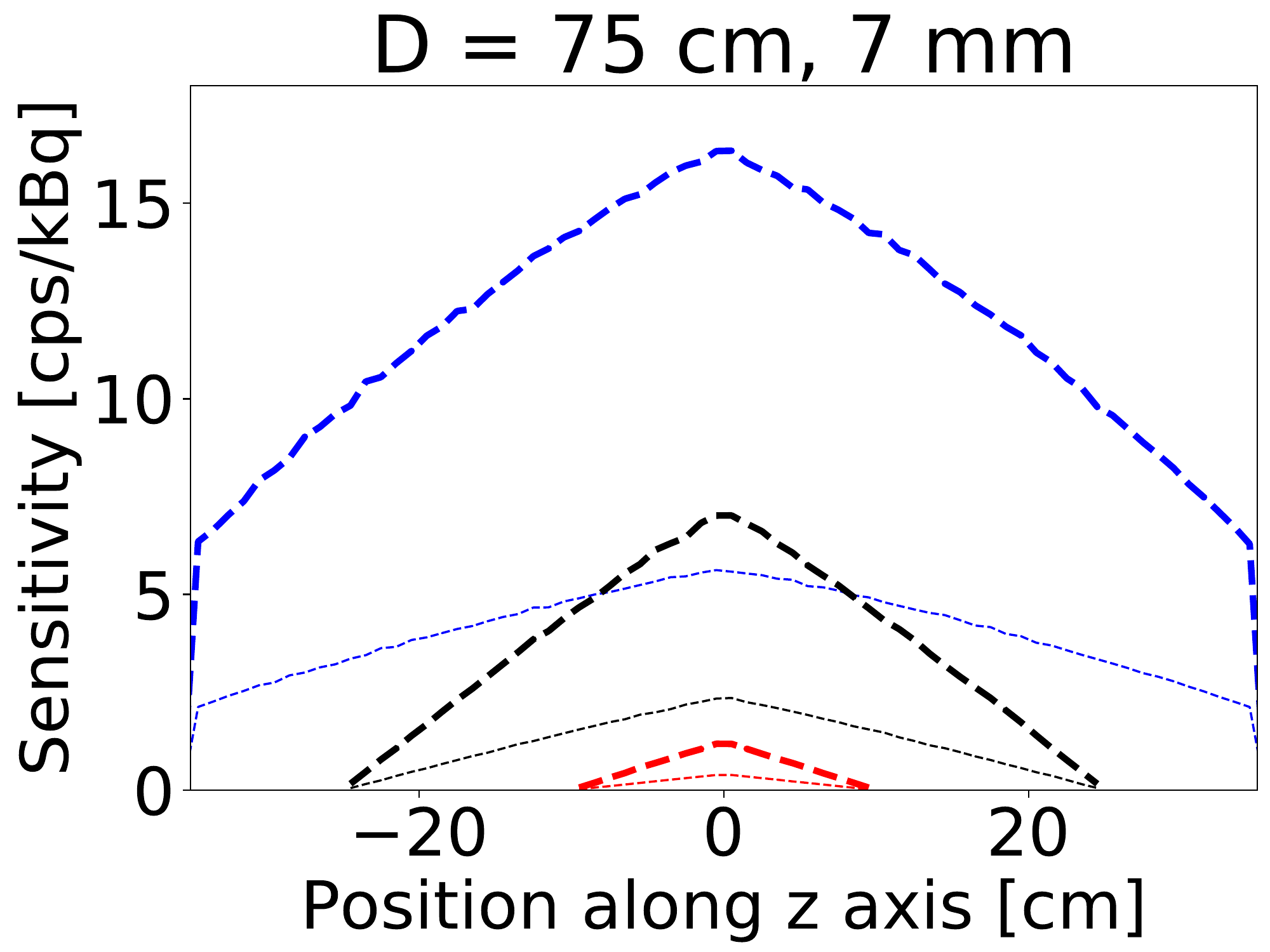}
\includegraphics[width=0.325\textwidth]{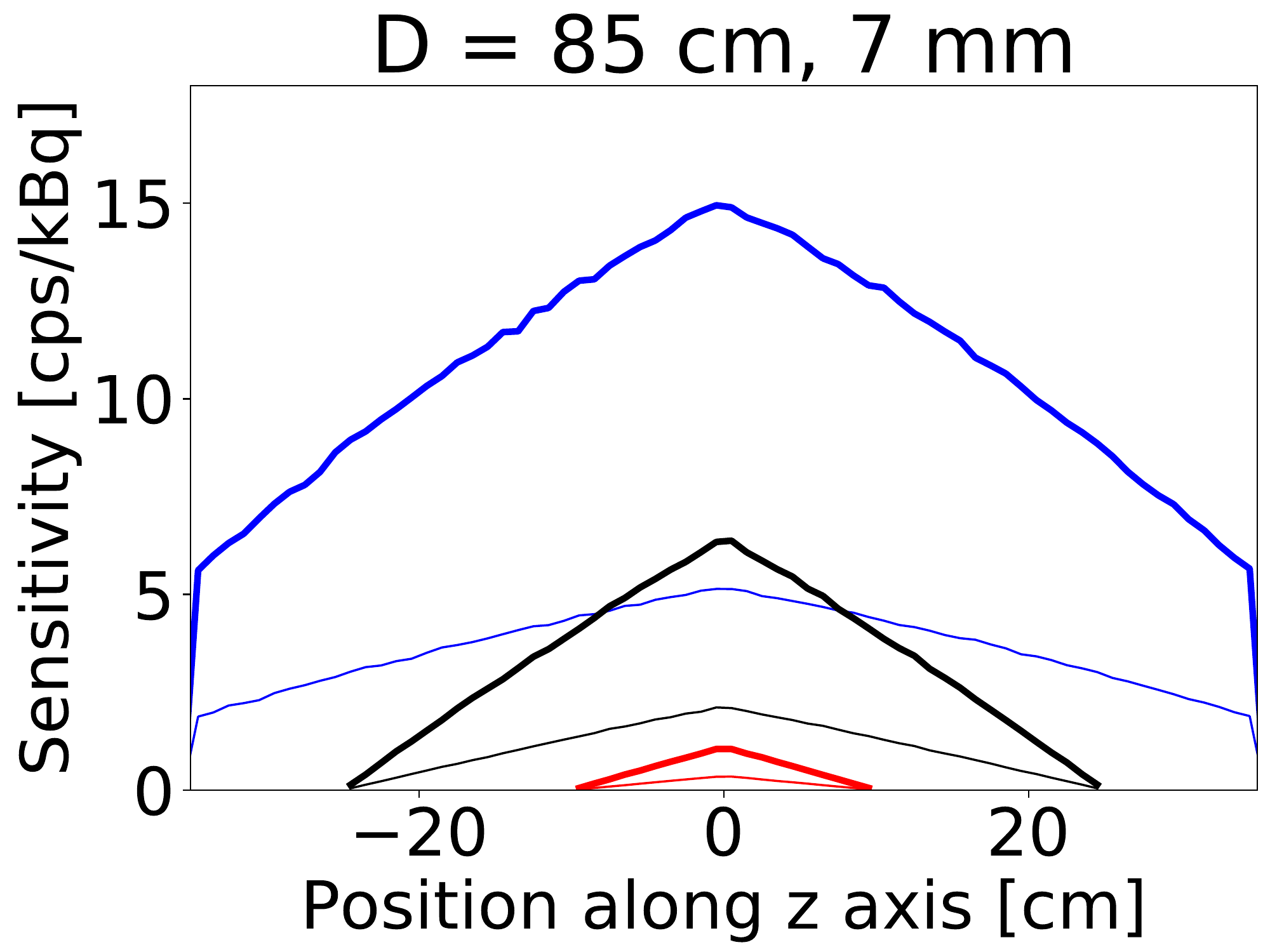}
\includegraphics[width=0.325\textwidth]{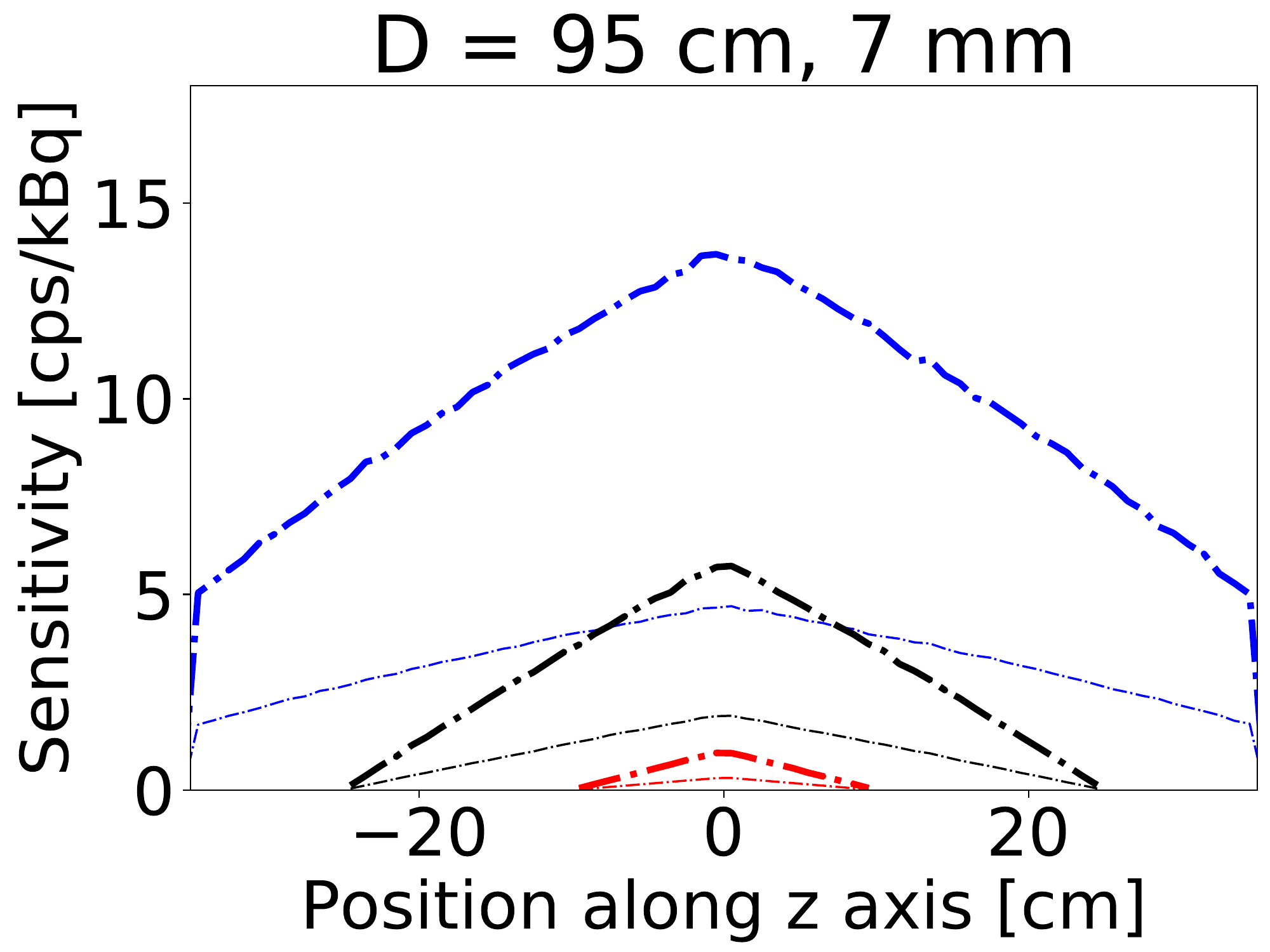}

\includegraphics[width=0.325\textwidth]{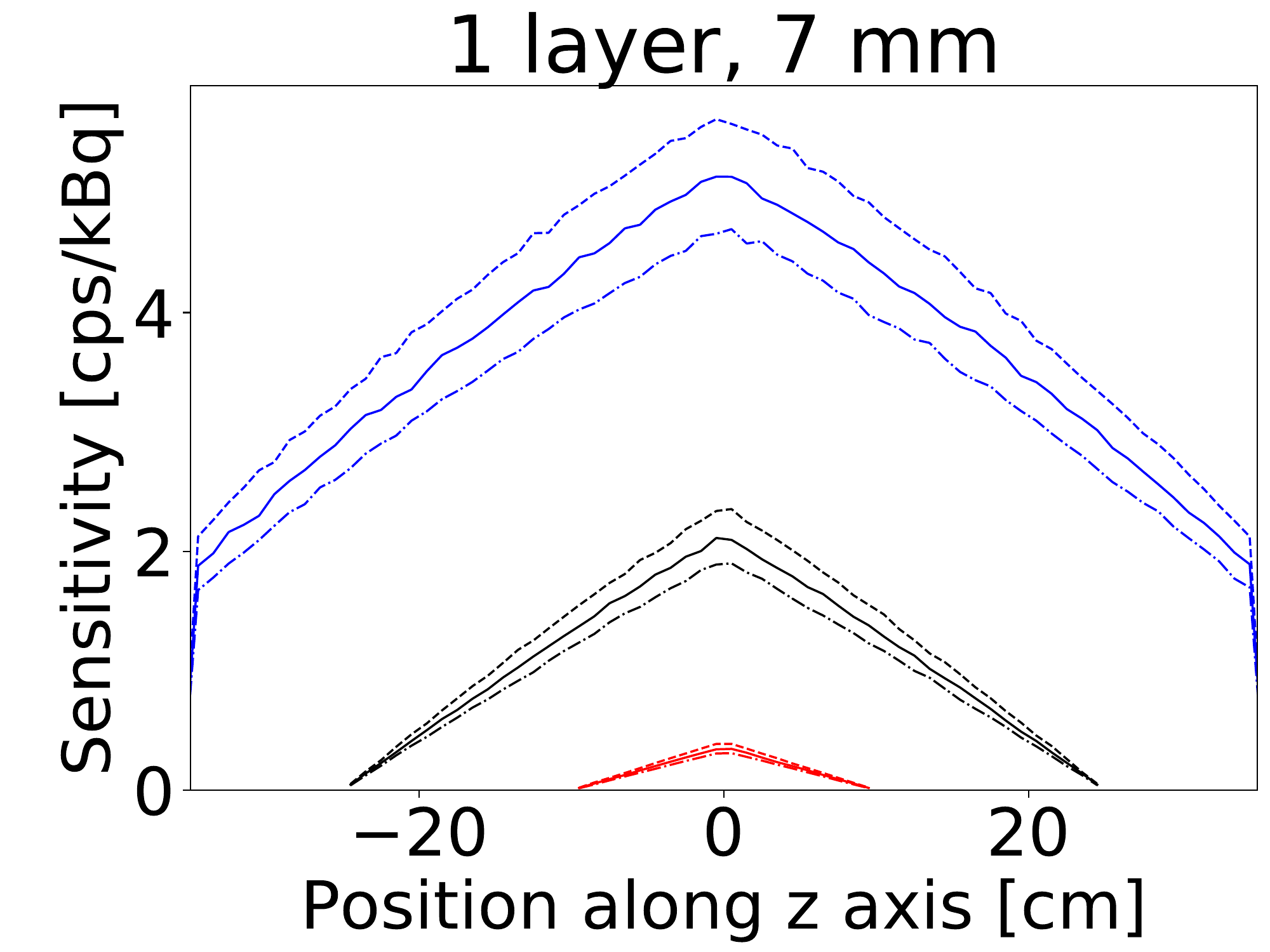}
\includegraphics[width=0.325\textwidth]{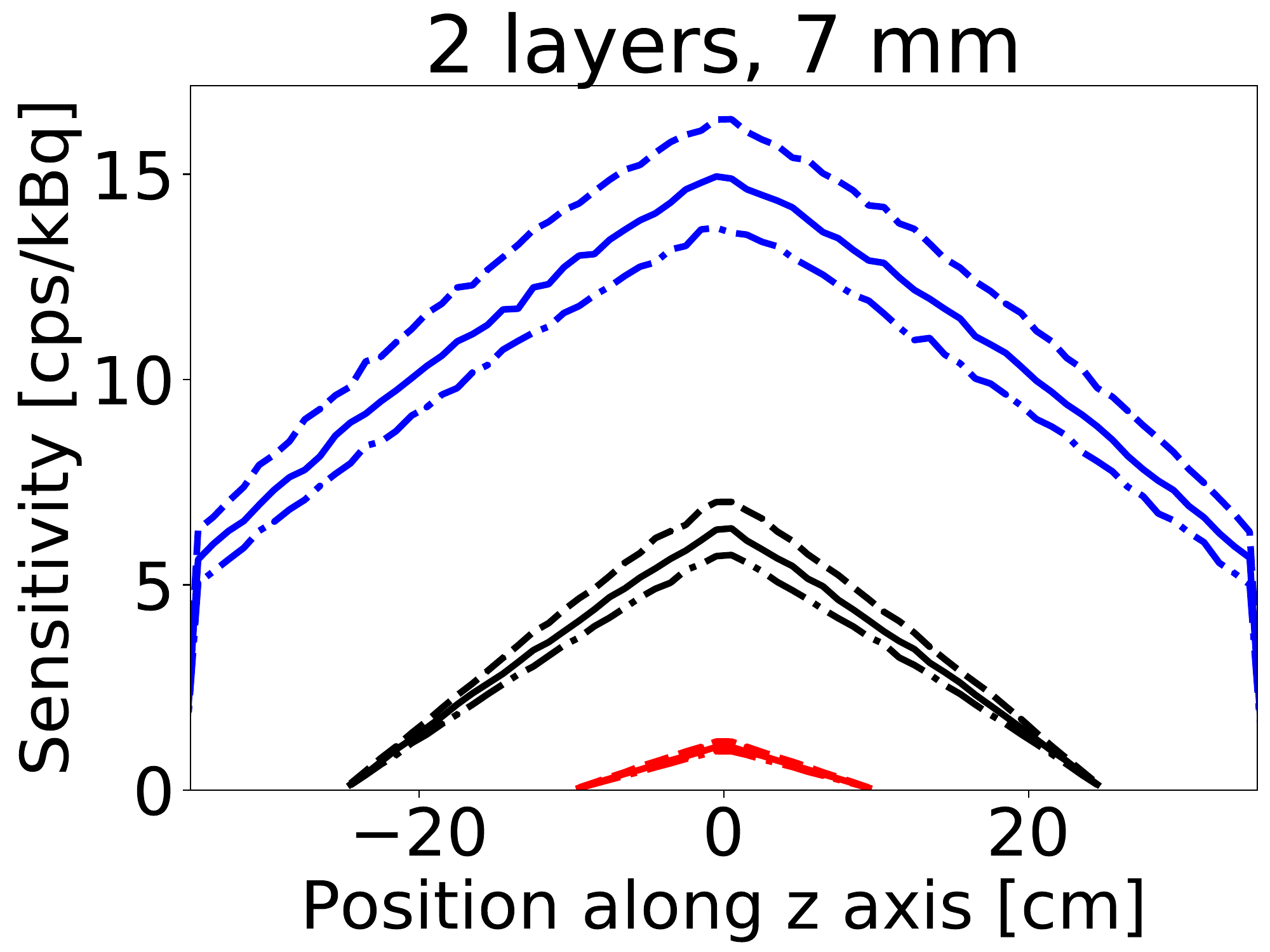}

\includegraphics[width=0.325\textwidth]{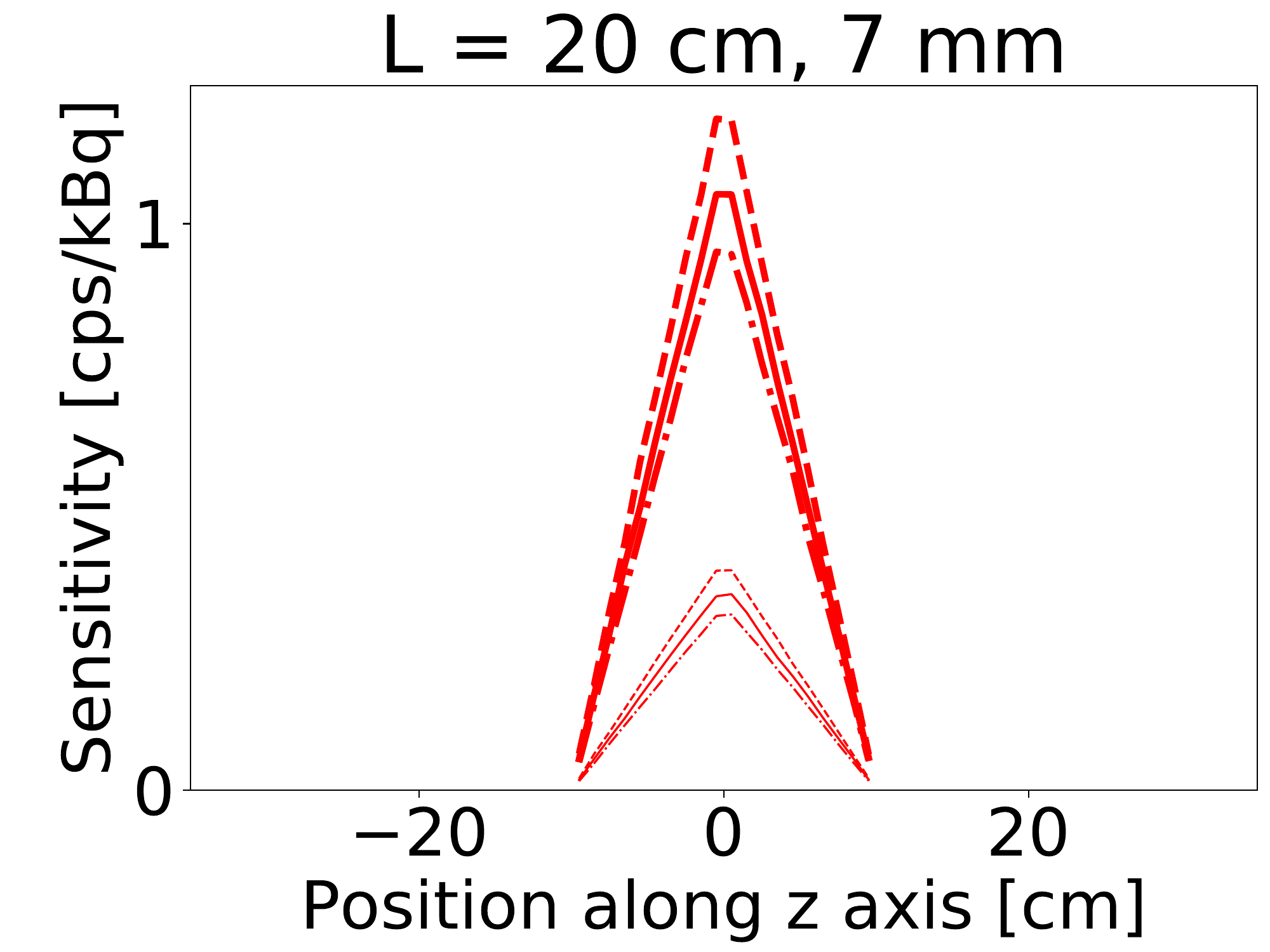}
\includegraphics[width=0.325\textwidth]{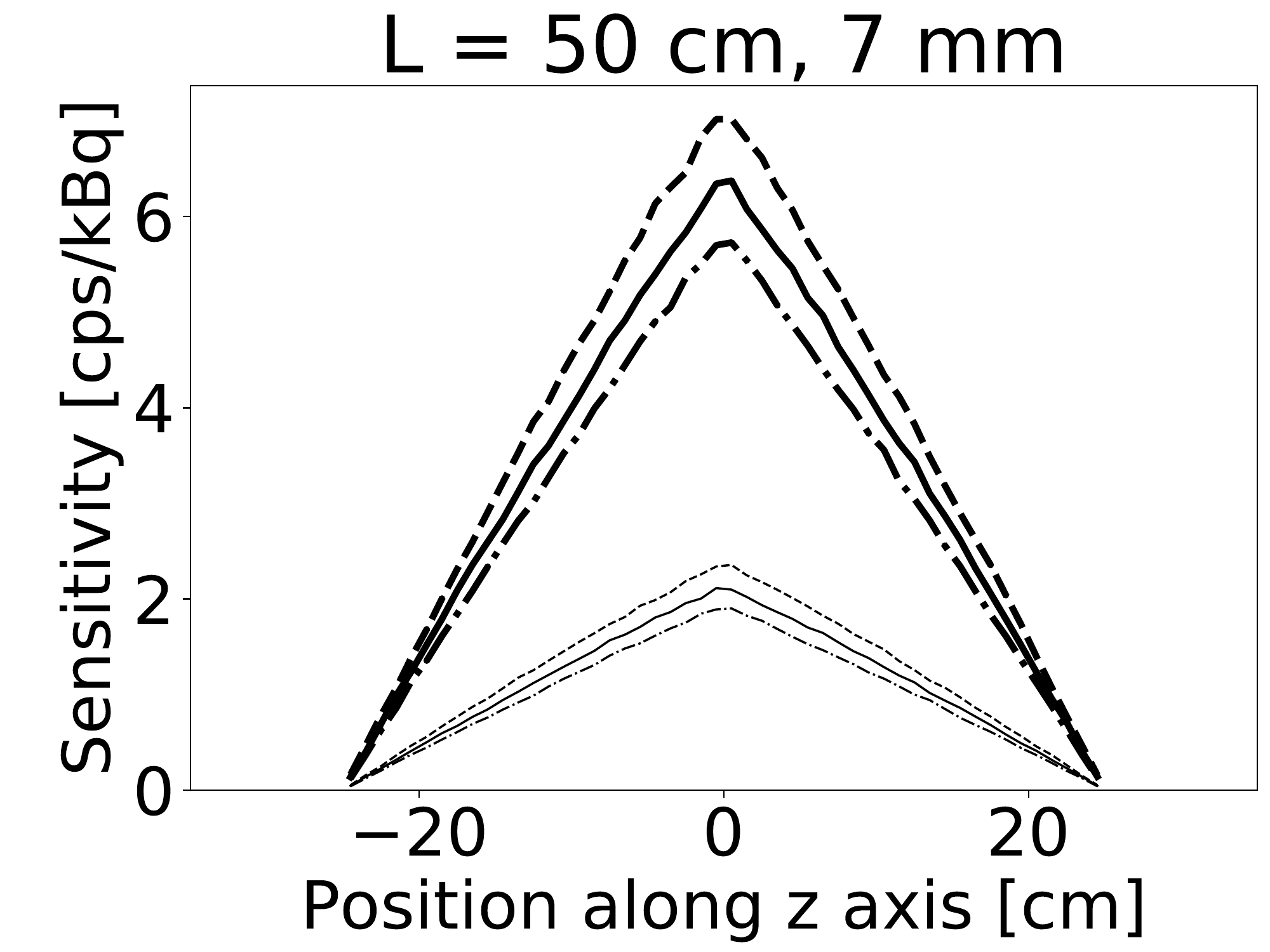}
\includegraphics[width=0.325\textwidth]{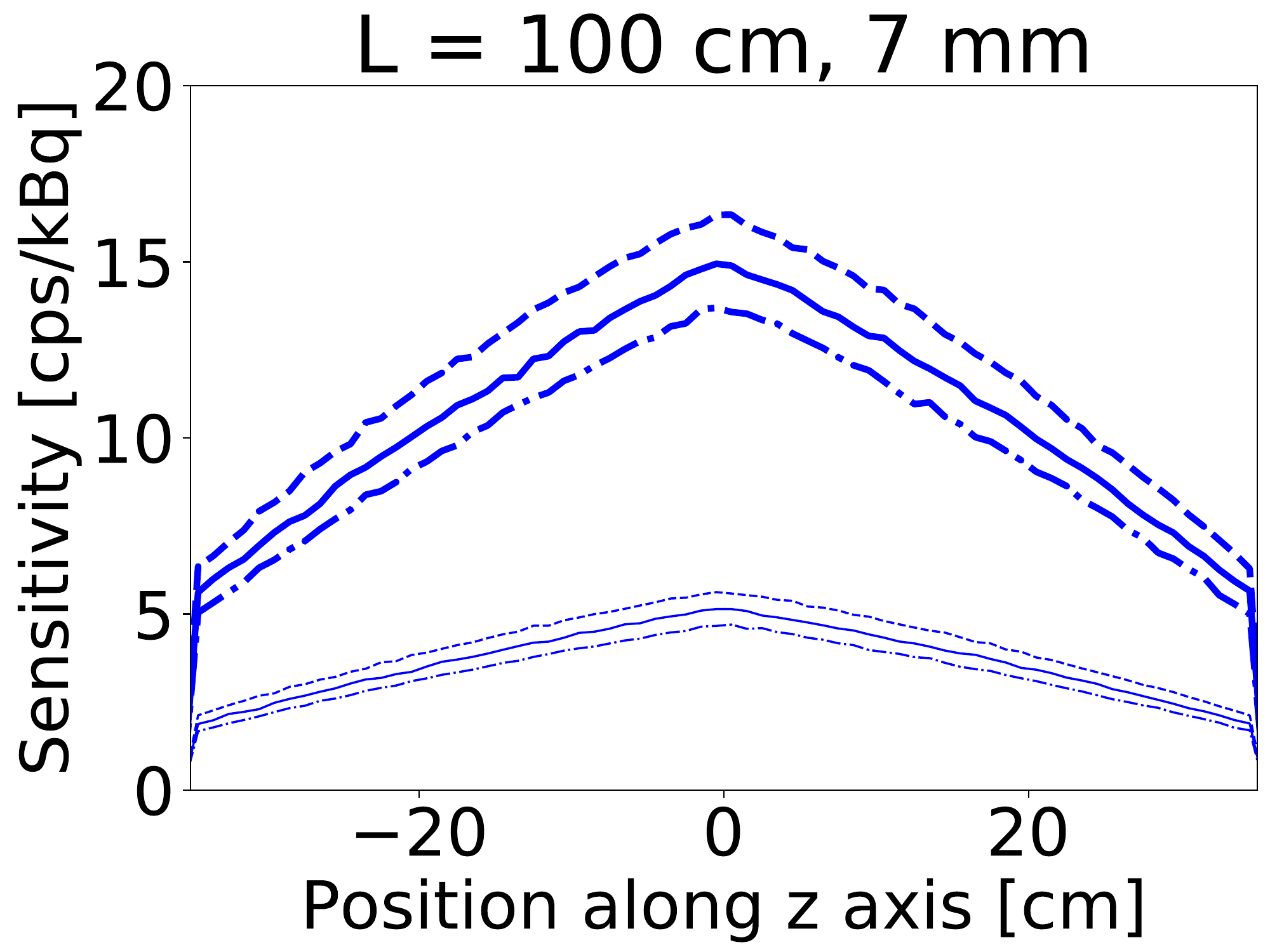}

\end{center}
\caption{
Sensitivity profiles grouped by the diameter of the detecting chamber (1st row), number of layers (2nd row) and lengths of the scintillators (3rd row). Legend: D = 75 cm (dashed - -), D = 85 cm (solid -), D = 95 cm (dashed-dotted -.-), 1 layer (thin lines), 2 layers (thick lines), L = 20 cm (red), L = 50 cm (black), L = 100 cm (blue).
}
\label{sensitivityProfiles}
\end{figure}

Profiles of sensitivity are presented in Fig. \ref{sensitivityProfiles}.
Since the length of the phantom was equal to 70~cm, the profiles are presented only for positions in range between -35~cm and 35~cm (along the axis of the scanner).

One can see, that values of sensitivities obtained for single- and double-layer geometries with the same diameters and lengths of strips, differ about 3~times.
At first moment, one could suppose that the sensitivity for the double-layer geometry would be almost 4~times bigger than for the corresponding single-layer geometry.
However, there are two reasons, that the ratio between the corresponding sensitivities is about~3.
Firstly, probability of detection of the single gamma photon with two layers is about 1.82~times bigger than with one layer (the efficiency of the registration of the 511~keV photon in the first layer is equal to about 17.8\%).
It means that the probability of registration of the coincidence with double-layer geometry would be about $1.82^2 = 3.31$ times higher than for single-layer geometry with the same strips length.
Secondly, not all true coincidences are properly registered due to a~fact that if beside the scatterings forming the true coincidence, the additional (scattered or accidental) interaction occurs during the time window, the event is rejected.
The effect is the stronger, the higher the detection efficiency (which grows with the number of layers and with the length of strips).
These both effects lead to the fact, that the improvement factor in terms of sensitivity is only about~3.

Adding the second layer of strips increases the detection efficiency.
However, the rate of the detector-scattered and accidental events grows faster than the rate of true coincidences.
This is because in case of two layers there is additional probability for interlayer scatterings which are not increasing rate of true coincidences but may increase the rate of scattered and accidental coincidences.

\FloatBarrier

\subsection{Spatial resolution}
\label{sr_section}

\begin{figure}[!htb]
\begin{center}
\includegraphics[width=250pt]{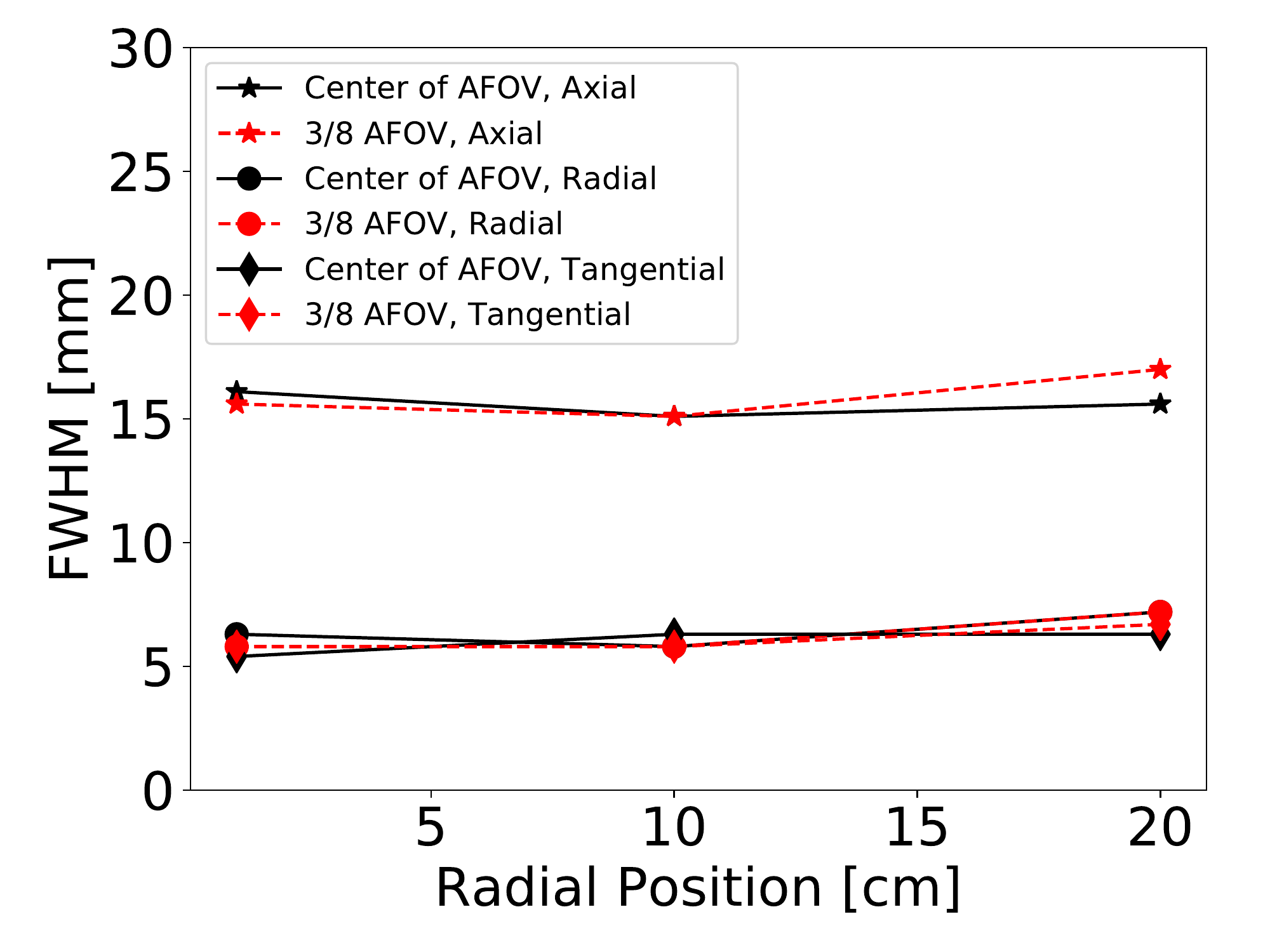}
\end{center}
\caption{
Spatial resolution in three directions - the radial, tangential and axial.
The geometry was fixed to the single layer chamber with the diameter of 85~cm and strips with length of 50~cm and thickness 7~mm.
Hit-time and hit-position resolutions were used as anticipated for the SiPM readout.
}
\label{SR_Axial_Position}
\end{figure}

Results were selected in order to show influence of each parameter on the final spatial resolution of the scanner.
Firstly, the influence of the axial position of the scanner was investigated.
The geometry was fixed to the single layer chamber with the diameter of 85~cm and strips with length of 50~cm and thickness equal to 7~mm.
Hit-time and hit-position resolutions were used as anticipated for the SiPM readout.
Results are presented in Fig.~\ref{SR_Axial_Position}.
All resolutions seem to be slightly dependent on radial position of the source.
There is also no visible difference between resolutions for sources placed in the centre of the AFOV and in 3/8 of the AFOV.

\begin{figure}[!htb]
\begin{center}
\includegraphics[width=250pt]{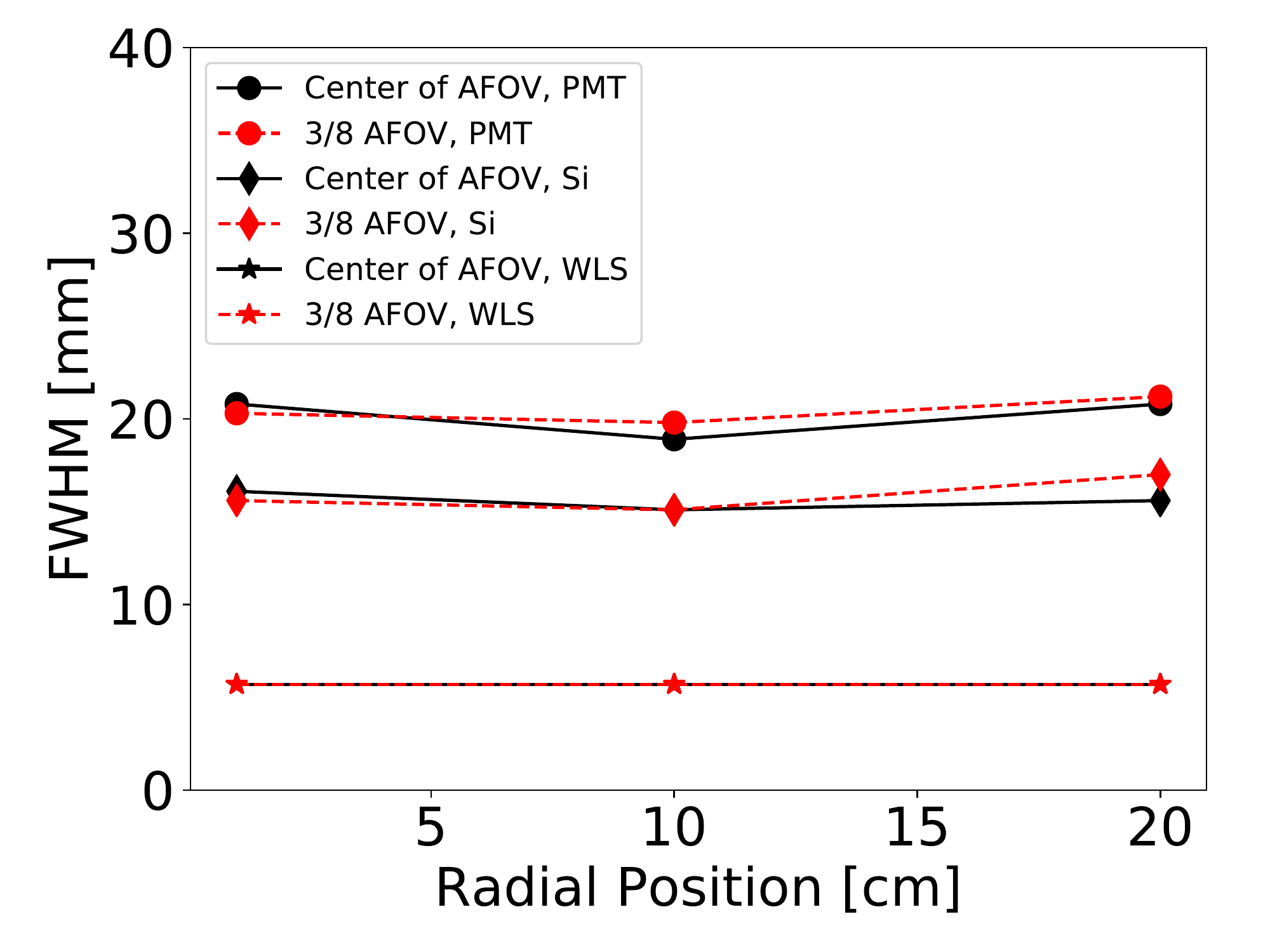}
\end{center}
\caption{
Axial resolution for different types of optical photon detectors attached to scintillator strips and two axial positions as a~function of radial position of the source.
Presented results were obtained for the single-layer geometry with the diameter of 85~cm and strips with length of 50~cm and thickness 7~mm.
Note, that the results for WLS at the centre (black stars) and at 3/8 AFOV (red stars) overlap.
}
\label{SR_Detector}
\end{figure}

Secondly, the type of the optical photon detector attached to the strip was taken into account (PMT, SiPM or WLS).
Geometry was fixed to the single layer chamber with the diameter of 85~cm and strips with length of 50~cm and thickness equal to 7~mm.
Results for different types of detectors are presented in Fig. \ref{SR_Detector}.
It is important to note, that only axial spatial resolution depends on the kind of the anticipated light readout.
Simulations confirmed intuition, viz that the smaller the experimental uncertainty, the better the axial resolution.
Additionally, they showed that if the number of collected coincidences is large enough, the spatial resolution barely depends on the position of the source along the axis of the scanner.

\begin{figure}[!htb]
\begin{center}
\includegraphics[width=0.495\textwidth]{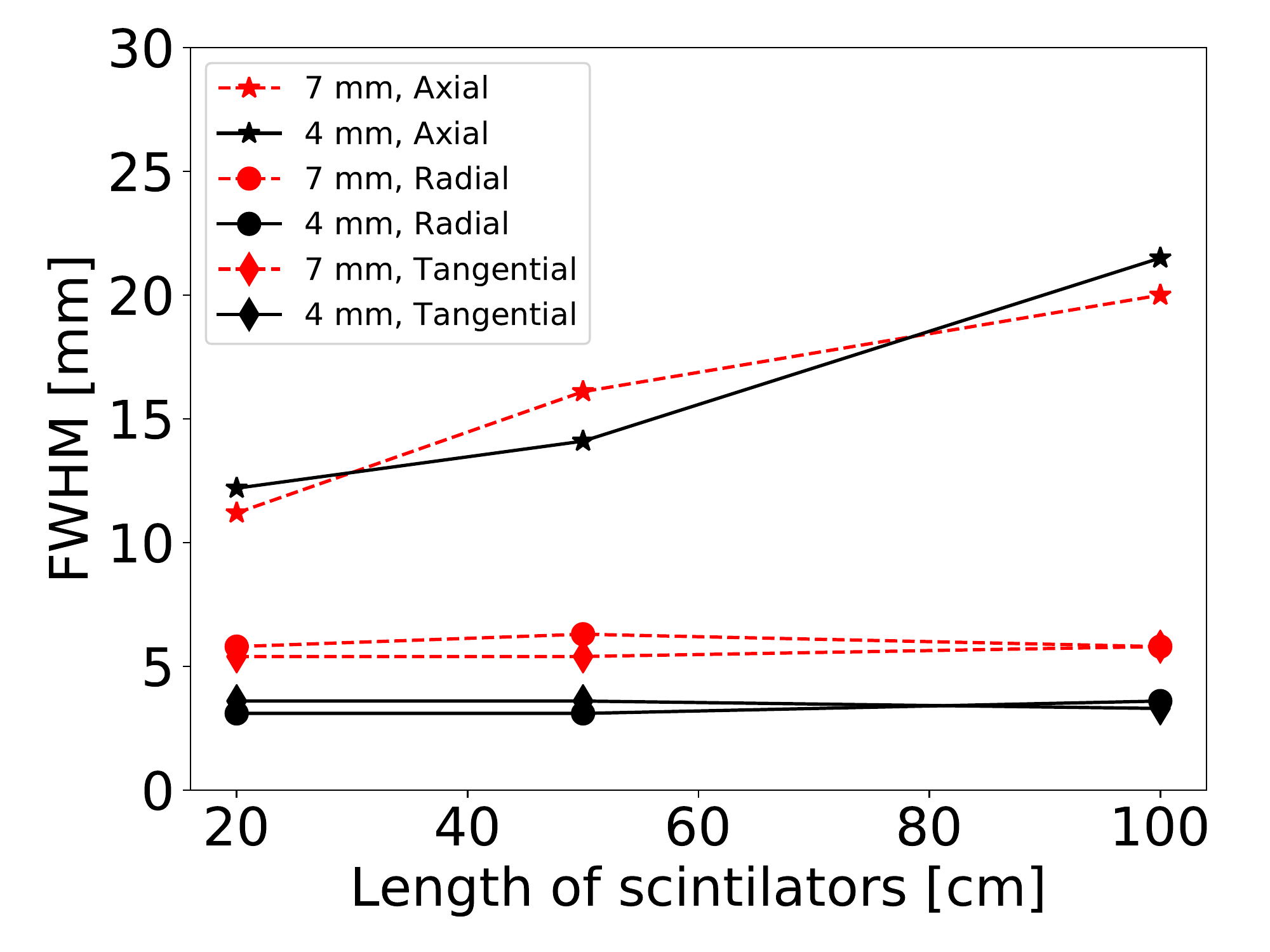}
\includegraphics[width=0.495\textwidth]{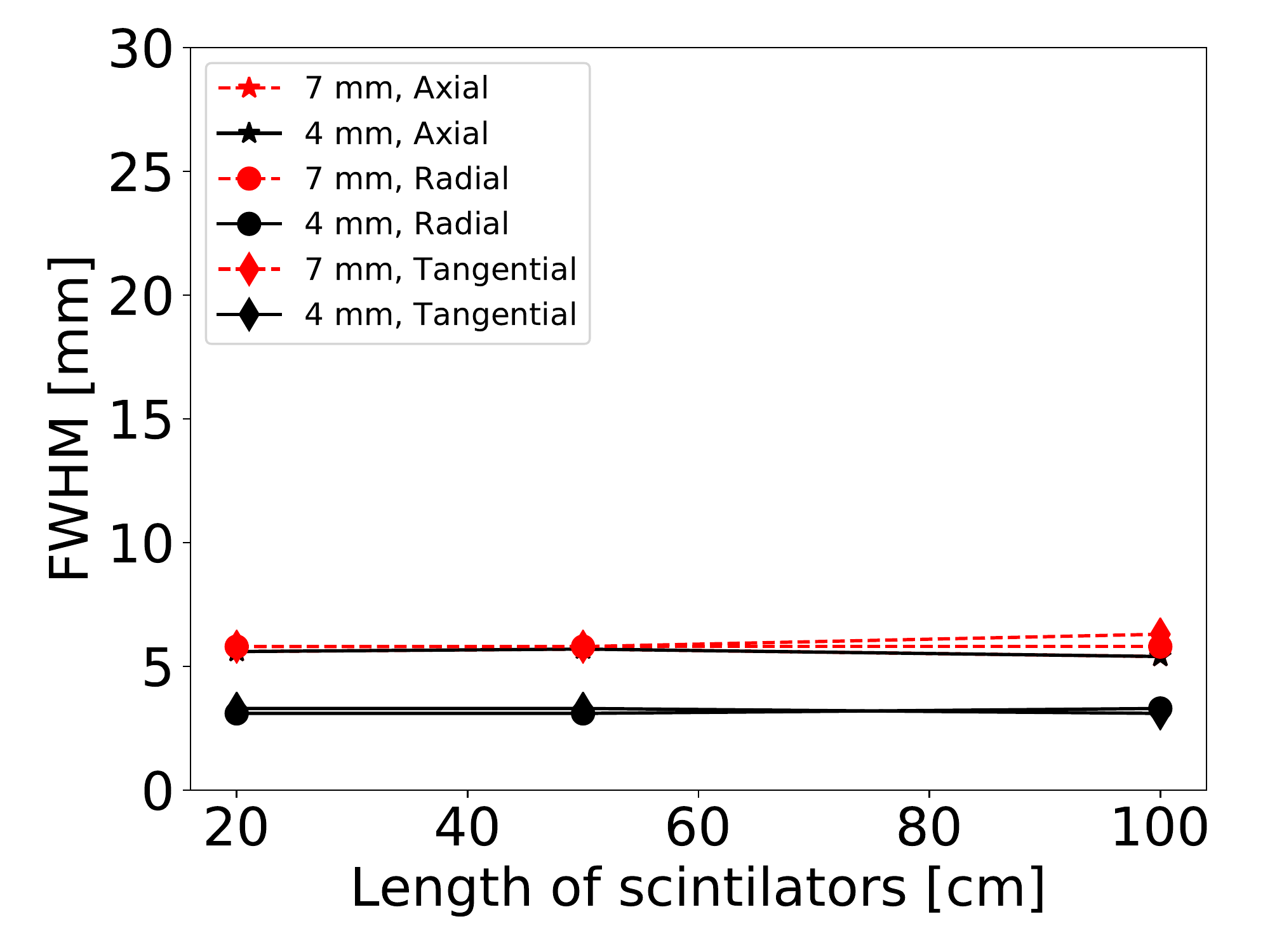}
\end{center}
\caption{
Spatial resolution for 3~lengths and 2~thicknesses of scintillator strips assuming readout from silicon photomultipliers (left) and silicon photomultipliers with the WLS strips (right).
}
\label{SR_D85_SiPM_01_00_00}
\end{figure}

Finally, the geometry of the single strip was taken into account.
Results for 3~lengths and 2~thicknesses of scintillator strips are presented in Fig. \ref{SR_D85_SiPM_01_00_00}.
Position of the source in each case was fixed to (1,0,0) cm (only 1~cm from the centre of the tomograph).
Simulations showed that, as expected in the case of the J-PET without WLS readout, the axial resolution worsens proportionally to the length of the strip (left panel of Fig.~\ref{SR_D85_SiPM_01_00_00}).
However, it improves and remains constant along the scanner in the case of the readout with the WLS layer (right panel of Fig.~\ref{SR_D85_SiPM_01_00_00}).
Because of the axial symmetry of the detecting chamber, radial and tangential resolutions are independent of the length of the strip. 
For each type of the resolution, better results are obtained for thinner strips (4~mm), e.g. for the radial and tangential resolution, they are reduced twice with respect to results obtained with strips of 7~mm thickness.
For the axial resolution there is a~slight difference between geometries with 4~mm and 7~mm thick strips.

\begin{figure}[!htb]
\begin{center}
\includegraphics[width=\textwidth]{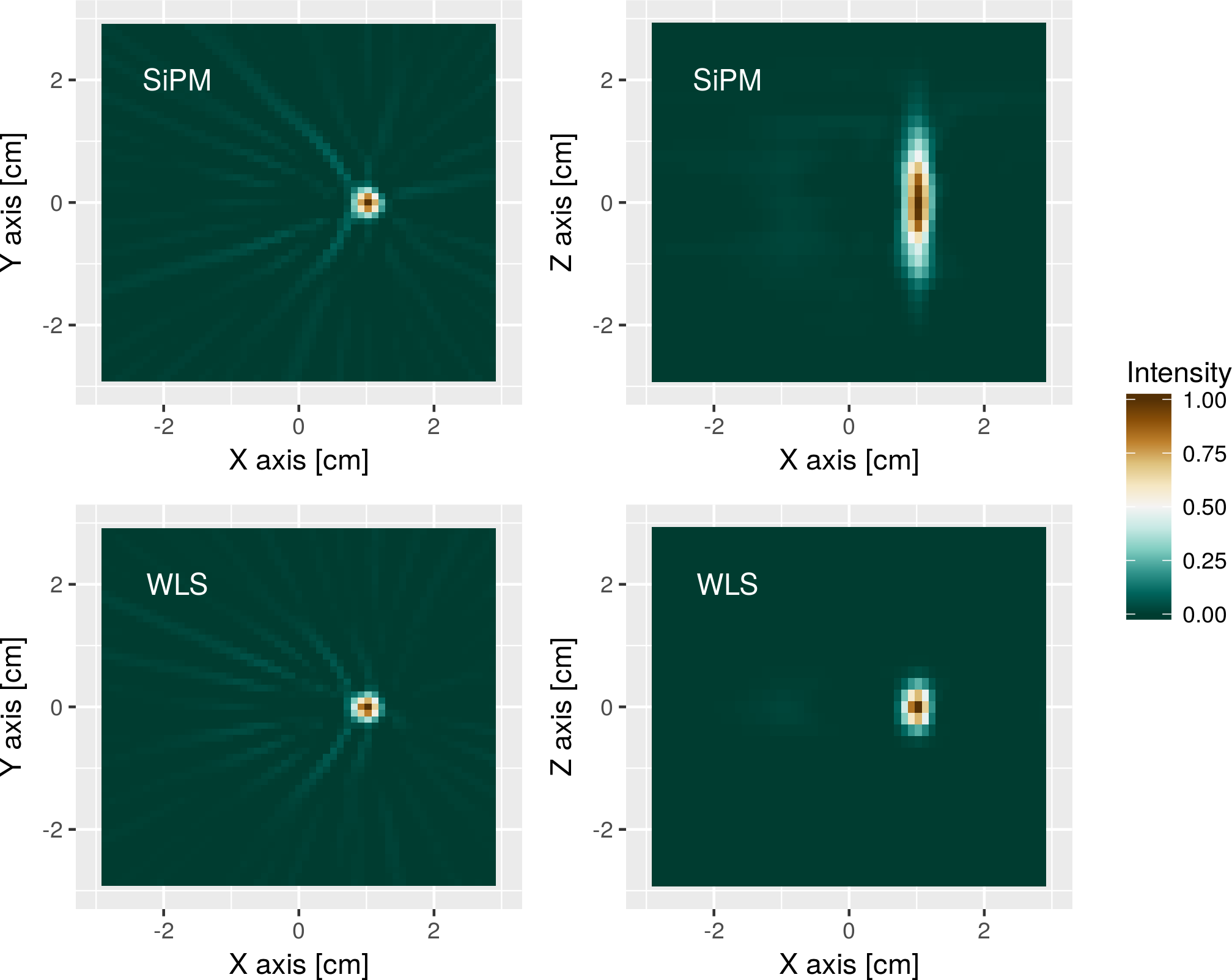}
\end{center}
\caption{
Example reconstruction of the source placed in the central position [(1,0,0) cm] of the detecting chamber.
The geometry consisted of the single layer chamber with the diameter of 85 cm and strips with length of 50~cm and thickness 4~mm.
Silicon photomultipliers (SiPM) were used as photodetectors in upper images, WLS strips were used in bottom images.
Left column corresponds to the cross-section perpendicular to the axis, right column to the cross-section along the axis of the scanner.
}
\label{example_recontruction}
\end{figure}

An example of reconstruction of the source placed in the centre of the tomograph may be seen in Fig. \ref{example_recontruction} (geometry was fixed to the single layer chamber with the diameter of 85~cm and strips with length of 50~cm and thickness equal to 4~mm; silicon photomultipliers were used as photodetectors).

\FloatBarrier

\subsection{Scatter fraction}
\label{sf_section}

Results obtained using the method based on the sinogram analysis are presented in Tab.~\ref{sf_results_sinograms}.
The scatter fraction was also calculated using the true Monte Carlo information about types of coincidences in each event.
Results for such calculations are presented in Tab.~\ref{sf_results_counters}.
Values obtained for these six geometries are consistent with the value calculated in previous studies for one layer 384~strip geometry \cite{Kowalski2016}.
The scatter fraction calculated at sinogram analysis is smaller than at the identification of events based on the information from Monte-Carlo simulations.
A~similar effect was reported in \cite{Yang2015}.


\begin{table}[!htb]
\begin{center}
\footnotesize
\begin{tabular}{|c|c|c|c|}
\hline
	\textbf{Nr of layers} &
	\textbf{L = 20 cm} &
	\textbf{L = 50 cm} &
	\textbf{L = 100 cm} \\
\hline
	1 &
	36.0\% &
	35.8\% &
	34.8\% \\
\hline
	2 &
	35.1\% &
	35.6\% &
	34.7\% \\
\hline
\end{tabular}
\caption{Scatter fraction for six geometries of the J-PET scanner calculated using the method based on sinograms analysis}
\label{sf_results_sinograms}
\end{center}
\end{table}


\begin{table}[!htb]
\begin{center}
\footnotesize
\begin{tabular}{|c|c|c|c|}
\hline
	\textbf{Nr of layers} &
	\textbf{L = 20 cm} &
	\textbf{L = 50 cm} &
	\textbf{L = 100 cm} \\
\hline
	1 &
	49.1\% &
	49.1\% &
	47.6\% \\
\hline
	2 &
	51.1\% &
	51.2\% &
	50.1\% \\
\hline
\end{tabular}
\caption{Scatter fraction for six geometries of the J-PET scanner calculated using the method based on true Monte Carlo}
\label{sf_results_counters}
\end{center}
\end{table}


\FloatBarrier

\subsection{Noise equivalent count rate}

\begin{figure}[!htb]
\begin{center}
\includegraphics[width=0.49\textwidth]{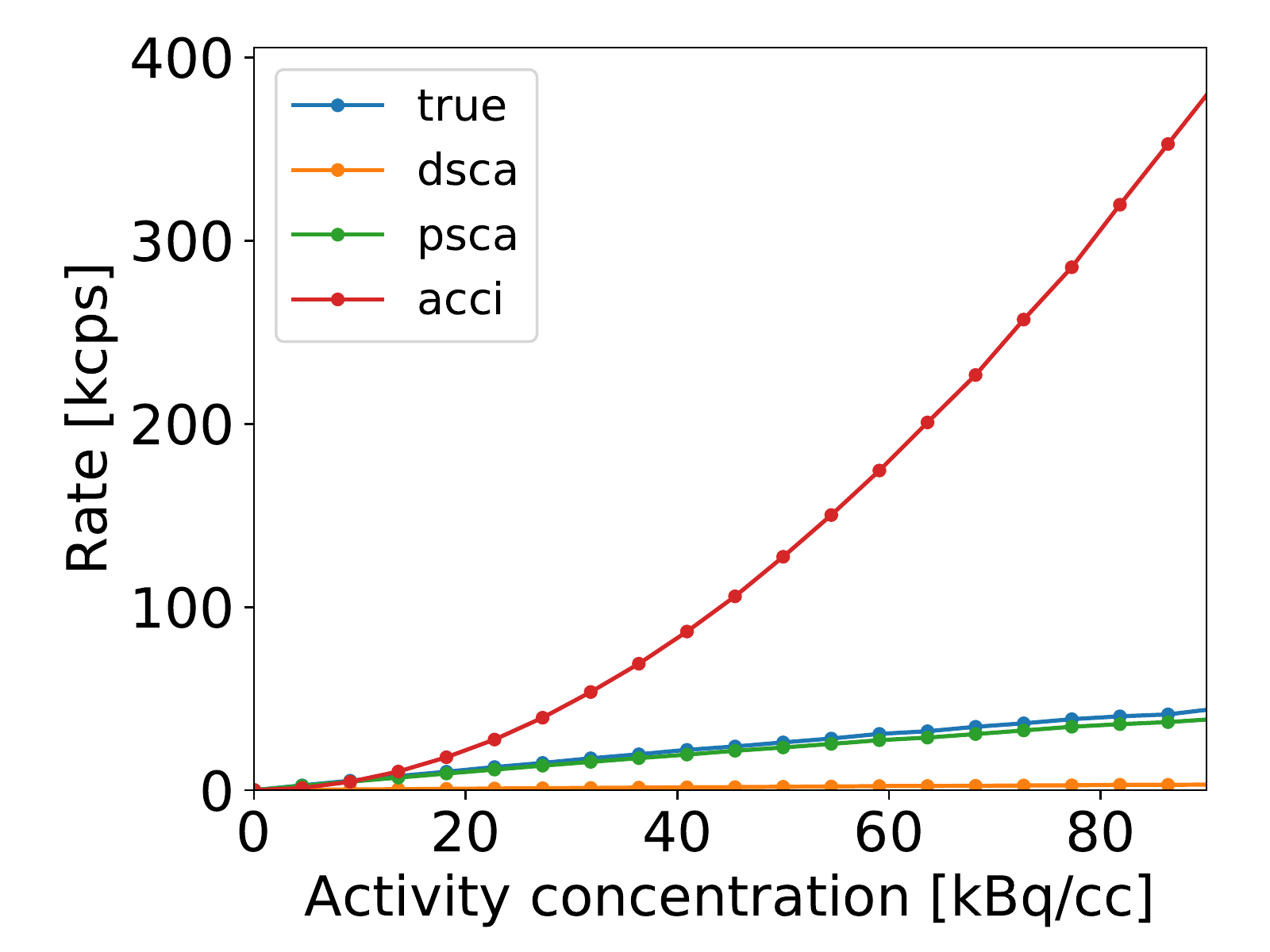}
\includegraphics[width=0.49\textwidth]{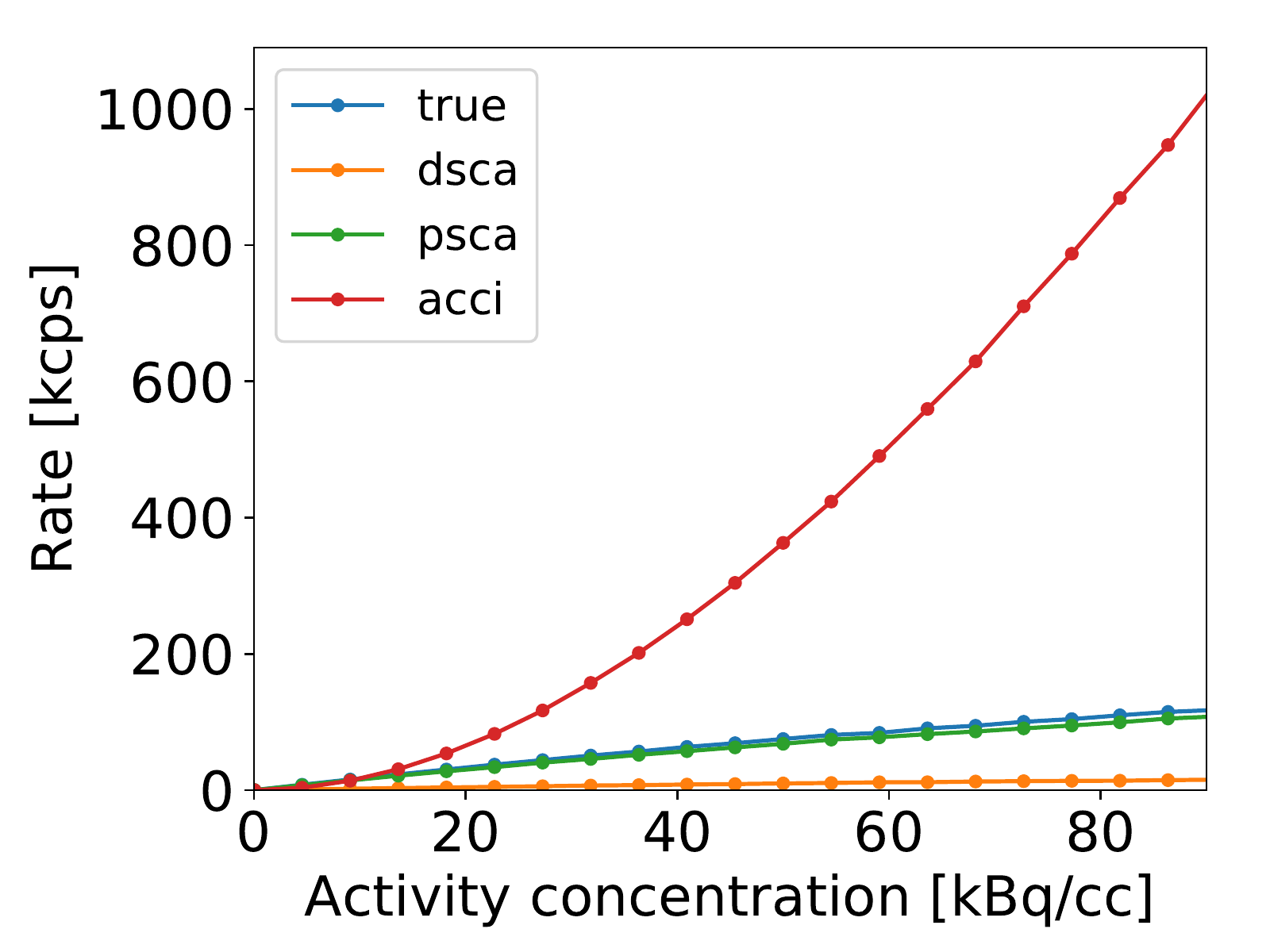}
\includegraphics[width=0.49\textwidth]{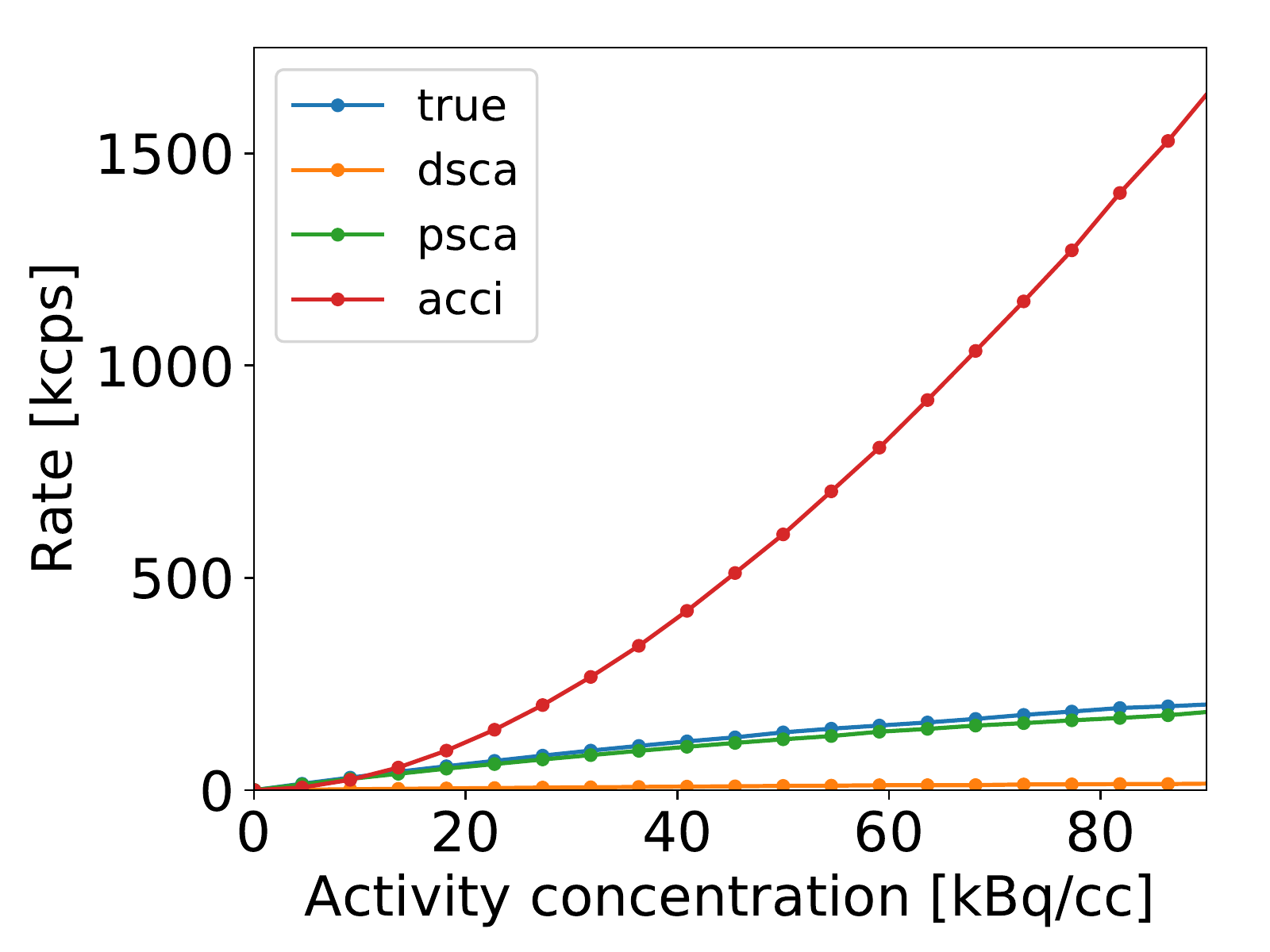}
\includegraphics[width=0.49\textwidth]{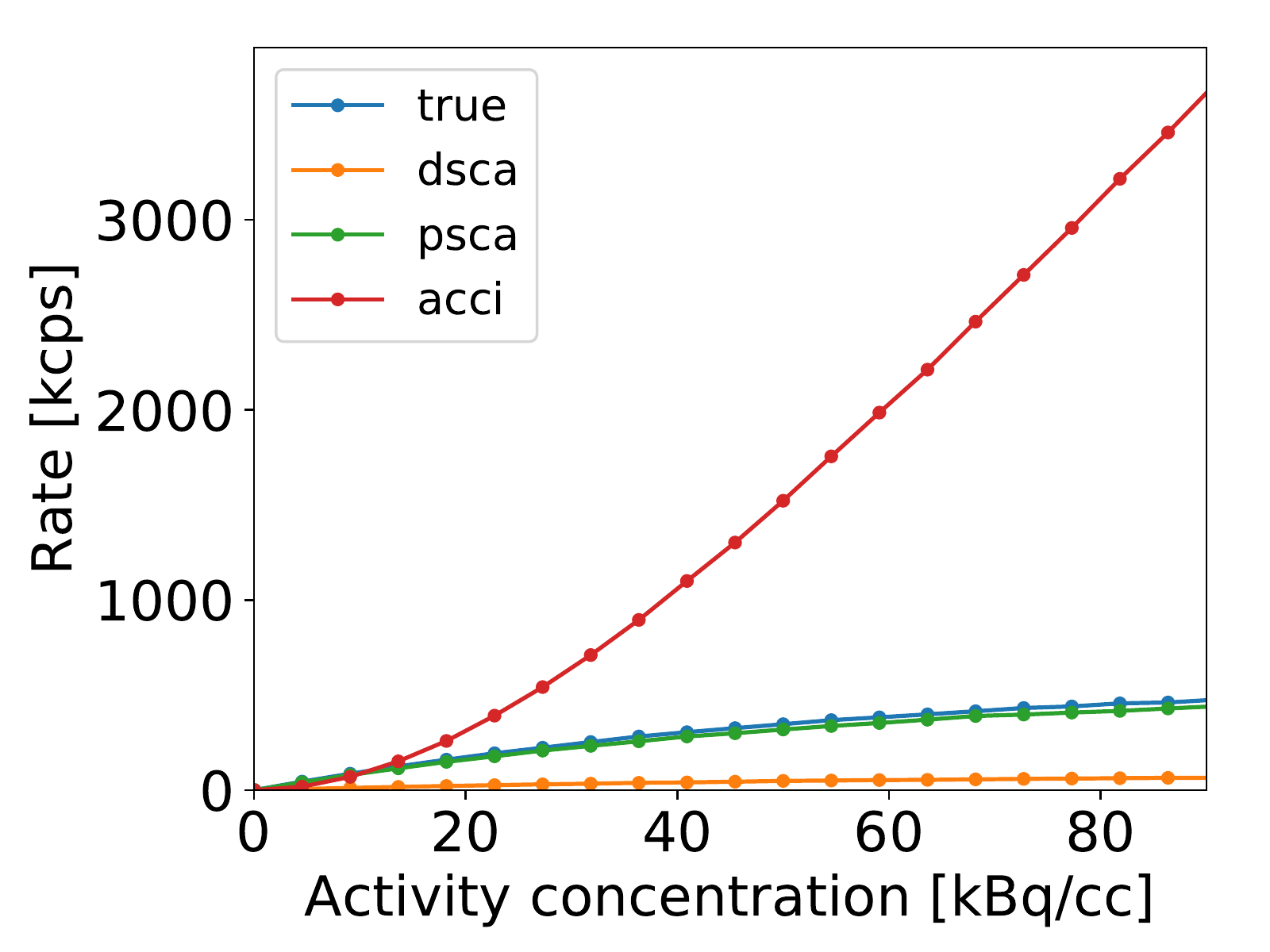}
\includegraphics[width=0.49\textwidth]{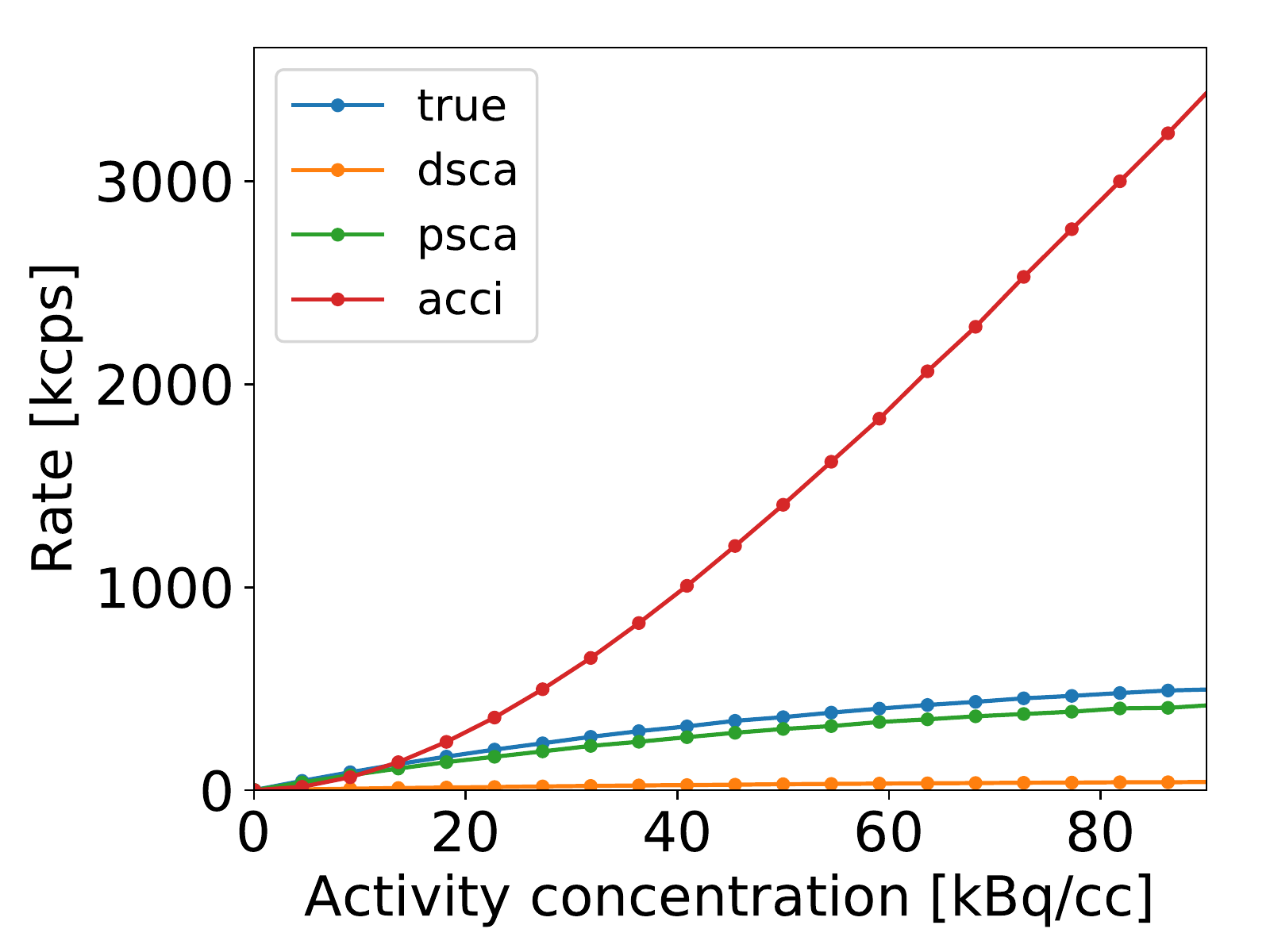}
\includegraphics[width=0.49\textwidth]{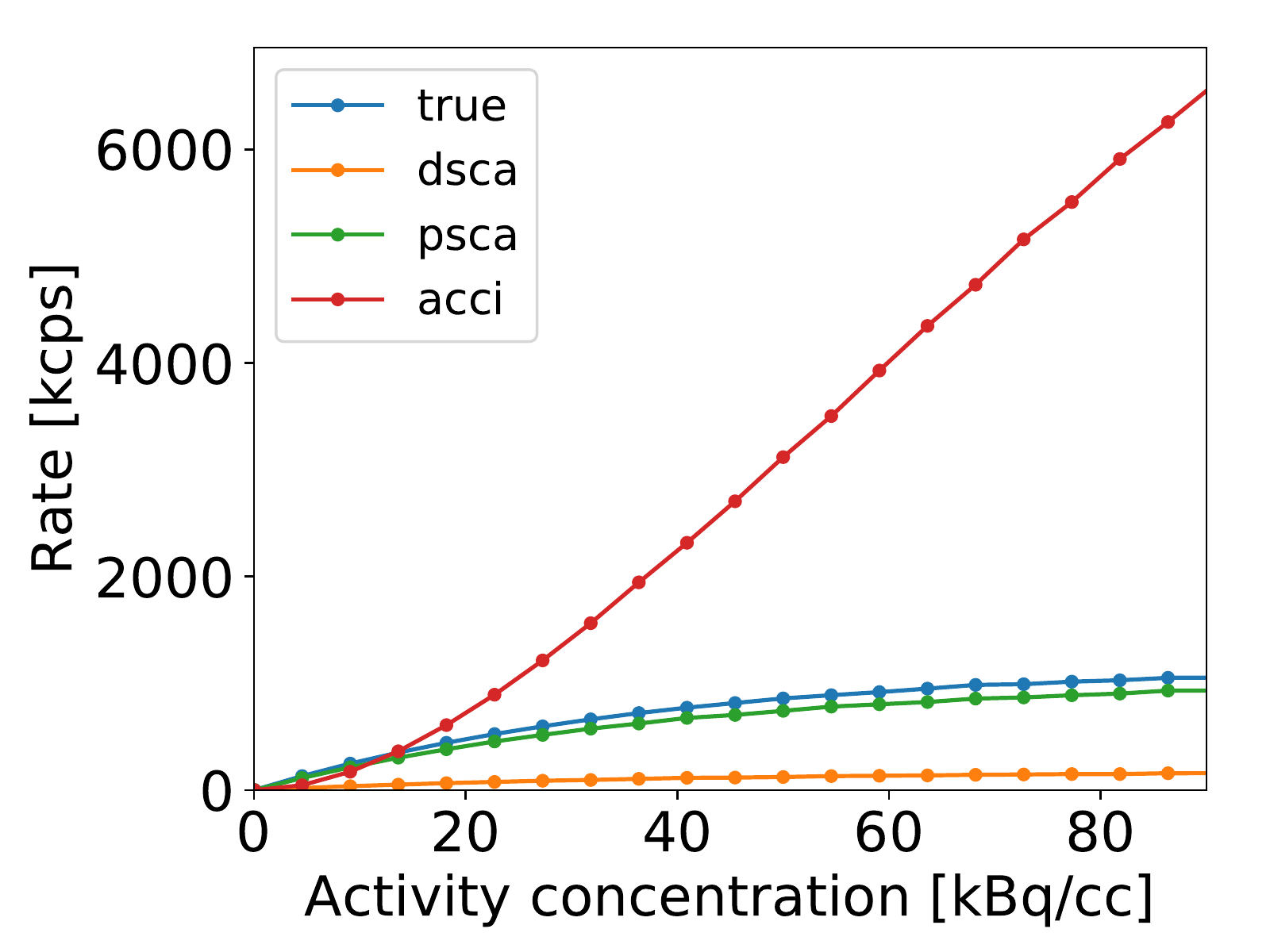}
\end{center}
\caption{
Count rates of different types of coincidences for 6~geometries of the J-PET scanner with diameter D~of 85~cm (dsca - detector-scattered, psca - phantom-scattered, acci - accidental coincidences).
Left and right panel show results obtained for 1- and 2-layer detectors, respectively.
In the first row of the figure, there are rates for 20~cm scintillators, in second row for 50~cm scintillators and in the third row for 100~cm scintillator strips.
}
\label{ratios_of_coincidences_3ns}
\end{figure}

Firstly, rates of the true, scattered and accidental coincidences were calculated (Fig. \ref{ratios_of_coincidences_3ns}).
Simulations were performed for activity concentrations in the range between 0~and about 90~kBq/cc (activity of source put inside the phantom was in range between 1~MBq and 2000~MBq).

\begin{figure}[!htb]
\begin{center}
\includegraphics[width=0.49\textwidth]{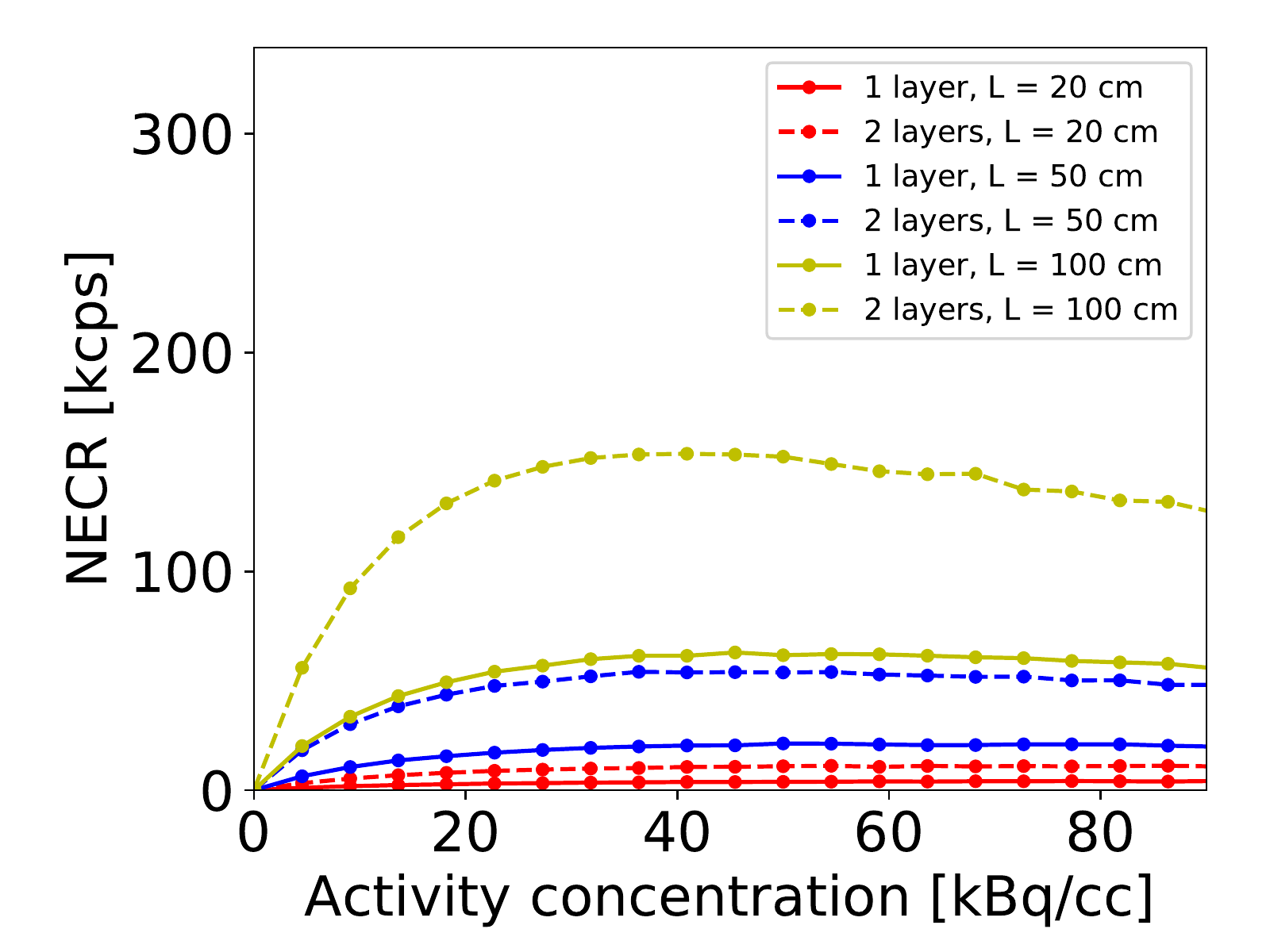}
\includegraphics[width=0.49\textwidth]{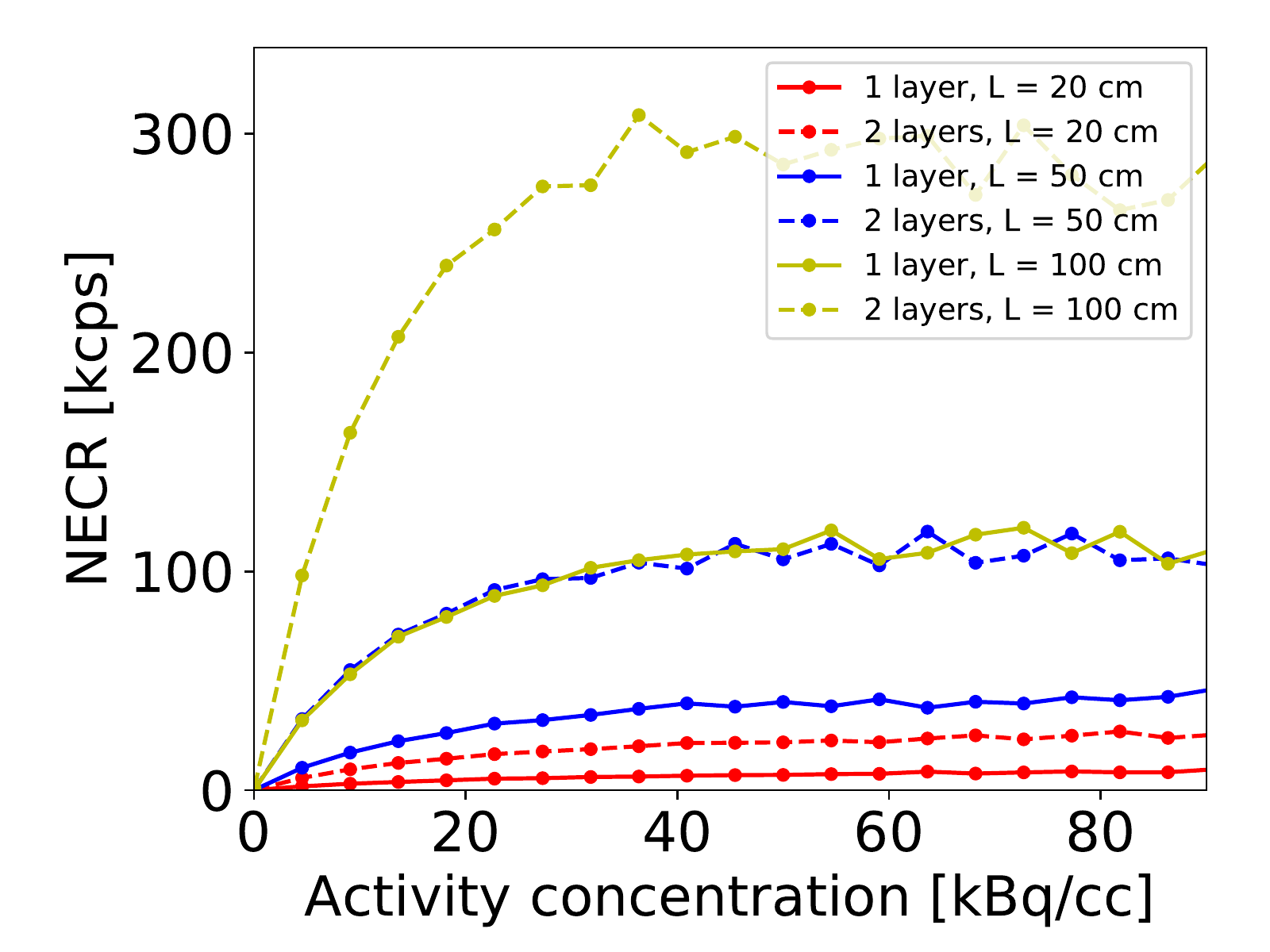}
\caption{
Noise equivalent count rate as a~function of activity concentration for six geometries of the J-PET scanner.
(left) NECR calculated using the true MC rates.
(right) NECR calculated using the method based on the sinograms analysis.
}
\label{necr}
\end{center}
\end{figure}

Secondly, the NECR characteristic was calculated (Fig. \ref{necr}).
As one can see, peaks for single layer tomographs are obtained for higher activity concentrations than for two layer scanners (for the same length of strips).
On the other hand, the longer the strips, the lower the value of activity concentration, for which the peak is obtained.
The best results were obtained for the geometry with two layers and 100~cm long strips.
For this geometry, the NECR peak was about 300~kcps for activity concentration of about 40~kBq/cc, for the method based on the sinograms analysis.
The method based on the true MC gives characteristics, which are about twice lower than in case of characteristics obtained using the sinograms analysis.
It is due to a usage of 12~cm radius cylindrical cut in the processing of sinograms \cite{NEMA, Yang2015}.


\section{Summary}
\label{section_summary}

Studies presented in this article cover the estimation and the analysis of the NEMA norms for the J-PET scanner.
Investigations were performed using the GATE software.
The spatial resolution, the scatter fraction, the NECR and the sensitivity were estimated according to the NEMA norm as a~function of the length of the tomograph, the number of the detection layers, diameter of the tomographic chamber and as a~function of the applied type of readout.

Firstly, the sensitivity profiles were estimated for three lengths of scintillator strips (20~cm, 50~cm, 100~cm), three diameters of detecting chamber (75~cm, 85~cm, 95~cm) and two thicknesses of scintillators (4~mm, 7~mm).
The sensitivity grows with the length of scintillators and with the number of layers (it is bigger for 2-layer geometries than for 1-layer geometries).
On the other hand, the larger the diameter of detecting chamber, the smaller the sensitivity.

For 2-layer geometry with diameter equal to 85 cm and with 50~cm scintillating strips, the sensitivity at the centre of the tomograph was about 6.3~cps/kBq while for 100~cm strips it exceeded 14.9~cps/kBq.
Sensitivity of the double-layer J-PET with 100~cm AFOV is in range of typical values of modern commercial PET scanners. For example for GE Discovery IQ, sensitivity measured at the centre and at 10 cm is 22.8 and 20.4 cps/kBq \cite{Llompart2017}, while for Philips Vereos it is about 21 cps/kBq \cite{PhilipsVereos}.

For each geometry configuration, the spatial resolution was calculated for six different positions of point source inside the detecting chamber.
All resolutions seem to be independent of the radial distance of source from the axis of the scanner.
While tangential and radial resolutions are independent of the length of the detecting strips, the axial resolution is the better the shorter strips are used.

The spatial resolution is strongly dependent on the type of readout simulated.
The shorter the CRT time and PSF(z), the better the resolution.
While in current J-PET prototype the vacuum tube photomultipliers are used, in next prototypes they will be replaced with the silicon photomultipliers and the WLS strips in order to reach higher class resolution.

Simulations showed that when the silicon photomultipliers were used (for geometry with diameter of 85~cm, with strips with length of 100~cm, thickness equal to 4~mm), the spatial resolution (PSF) in the centre of the scanner is about 3~mm (radial, tangential) and 20~mm (axial).
If an additional layer of WLS would be used to improve the readout, the axial resolution would be equal to 6~mm.
It is comparable to values of spatial resolutions for currently used commercial PET scanners.
For example GE Discovery IQ has spatial resolution in range of 4.2~mm at 1~cm to 8.5~mm at 20~cm \cite{Llompart2017}.
Similar resolutions (about 5~mm) are also obtained for Philips Gemini TF PET/CT \cite{PhilipsGemini, Surti2007}, Philips Vereos (4~mm) \cite{PhilipsVereos} and Siemens Biograph TM \cite{SiemensBiograph, Gonias2007}.

The scatter fraction was calculated for six geometries of the J-PET scanner with three different lengths of strips and one or two layers of scintillators.
The cylindrical NEMA phantom was used for simulations.
The values of scatter fraction are strongly dependent on the method used in order to calculate the characteristic.
If the method based on sinogram analysis was used, the scatter fraction was in the range of 34.7\% (2 layers, 100~cm strips) to 36.0\% (1 layer, 20~cm strips).
On the other hand, if method based on the true Monte Carlo was used, the scatter fraction was in the range of 50.2\% (2 layers, 50~cm strips) to 47.6\% (1 layer, 100~cm strips).

Obtained values of scatter fraction are similar to those computed and measured for commercial PET scanners.
For example, the GE Discovery has the scatter fraction between 21\% and 34\%, dependently on the used mode (2D or 3D) \cite{Teras2007}.
The newest model of GE tomograph - Discovery IQ has scatter fraction of 36.2\% \cite{Llompart2017}.

Next characteristic from the NEMA norm is the NECR dependency.
The NECR was obtained for geometries listed above for scatter fraction and for the same cylindrical phantom.
The activity concentration ranges from 0~to about 90~kBq/cc (Fig.~\ref{necr}).
The simulations showed that the longer the strips, the higher value of NECR peak and it is obtained for smaller value of activity concentration.
The smaller the activity concentration, the smaller dose deposited in the patient's body and shorter recuperation.

The best results of NECR were obtained for 2-layer geometry with 100~cm scintillator strips. The NECR peak for this geometry was equal to about 300~kcps and it was achieved at about 40~kBq/cc.
For 2-layer geometry with 50~cm strips the NECR peak of 110~kcps was reached at 63~kBq/cc.

For commercial PET scanners, the values of NECR peak are strongly dependent on the model of the tomograph.
For example, the GE~Discovery achieves the NECR peak of 84.9~kcps at 43.9 kBq/cc (2D) and 67.6~kcps at 12.1~kBq/cc (3D) \cite{Teras2007}, the GE Discovery IG achieves the NECR peak of 124 kcps at 9.1 kBq/cc \cite{Llompart2017} and the Siemens Biograph mCT achieves the 186 kcps at 30.1 kBq/cc \cite{Karlberg2016}.
It places the J-PET scanner in the range of typical values.
However, current studies show that there is a~place for a~significant improvement in reduction of the random background, which leads to higher values of NECR peaks and finally to images with higher qualities \cite{Oliver2016}.

The best spatial resolutions were obtained for the 4~mm thick strips and the SiPM readout with additional layer of the WLS strips.
Adding second cylindrical layer of strips seems to have slight influence on the spatial resolution and the scatter fraction, but it strongly improves the sensitivity and the NECR characteristics.
Increasing the diameter of the detecting chamber worsens the sensitivity of the scanner.
The longer the strips, the higher the sensitivity, the higher the NECR peak (which is obtained for smaller value of the activity concentration) and the smaller the scatter fraction.
On the other hand, without WLS strips, the axial resolution worsens with the growth of strips length.
In order to take into account all above figures of merit in process of projecting the J-PET prototype, the compromise must be found between the geometrical acceptance, the background, the image quality and the production cost.
It seems that from all tested geometries, the best results were obtained for the double-layer geometry built from strips with length of 100~cm and thickness of 4~mm, the diameter equal to 75~cm and the SiPM photomultipliers with the additional layer of the WLS strips used as readout.

The above studies confirmed that the PET scanner based on the plastic scintillator strips, may achieve results of the NEMA characteristics comparable to those obtained for commercially used PET scanners.
We believe that presented results may be improved.
For example, the method of events selection (section \ref{eventselection}) may be optimized.
Preliminary studies showed that there is strong dependence between the maximal number of additional hits in an event (with energies between 10~keV and 200~keV), value of the noise energy threshold and the values of presented characteristics.
The design of the scanner may be also optimized with shapes of strips and numbers of scintillator layers and it will be improved in future prototypes.


\section*{Acknowledgements}

We acknowledge technical and administrative support of A.~Heczko, M.~Kajetanowicz, W.~Migda{\l}, and the financial support by The Polish National Center for Research and Development through grant INNOTECH-K1/IN1/64/159174/NCBR/12, the Foundation for Polish Science through MPD and TEAM/2017-4/39 programmes, the National Science Centre through grants Nos. 2016/21/B/ST2/01222, 2017/25/N/NZ1/00861, the Ministry for Science and Higher Education through grants No. 6673/IA/SP/2016, 7150/E-338/SPUB/2017/1, the EU and MSHE Grant No. POIG.02.03.00-161 00-013/09, B.C.H. acknowledges support by the Austrian Science Fund (FWF-P26783).


\label{References}
\newcommand{\newblock}{}
\bibliographystyle{unsrtnat}
\bibliography{references,references_jpet} 

\end{document}